\documentclass[%
 reprint,
 amsmath,amssymb,
 aps,
]{revtex4-2}

\usepackage{graphicx}
\usepackage{dcolumn}
\usepackage{bm}
\usepackage{braket}
\usepackage{xcolor}
\usepackage[abs]{overpic}
\usepackage{soul}

\newcommand{\be}{\begin{eqnarray}}
\newcommand{\ee}{\end{eqnarray}}
\newcommand{\nn}{\nonumber}
\newcommand{\bea}{\begin{eqnarray}}
\newcommand{\eea}{\end{eqnarray}}

\renewcommand{\vec}[1]{{\bm #1}}

\begin{document}

\preprint{APS/123-QED}

\title{Valley-Controlled Viscosity of Two-Dimensional Dirac Fluids}

\author{Alexey Ermakov}
 \email{Alexey.Ermakov@manchester.ac.uk}
\author{Alessandro Principi}%
 \email{Alessandro.Principi@manchester.ac.uk}
\affiliation{%
 Department of Physics and Astronomy, 
 University of Manchester, Oxford Road, M13 9PL Manchester, United Kingdom
}%

\date{\today}

\begin{abstract}
Motivated by recent experiments in weakly hybridized small-angle twisted bilayer graphene, we investigate how valley imbalance affects the viscosity of two-dimensional Dirac fluids. We show that shifting the two low-energy Dirac cones relative to one another provides a direct knob to control the viscosity of the electron fluid. As the splitting is increased, the system passes through distinct transport regimes associated with valley depletion, charge-neutrality crossover, and the onset of electron-hole scattering, producing a pronounced nonmonotonic response.
To place this result in context, we also analyze the viscosity in monolayer graphene (MLG) and two-dimensional electron gas (2DEG). 
We show that, due to the strong dependence of its inertial mass density on temperature, the kinematic viscosity of MLG is a monotonically decreasing function of temperature.
Our results identify valley control as a route to tuning hydrodynamic transport in Dirac materials and clarify the interplay between band structure, scattering phase space, and screening in setting the viscous response.

\end{abstract}

\maketitle

\section{Introduction}

The hydrodynamic regime of electronic transport emerges when electron-electron collisions dominate over momentum-relaxing processes due to disorder, phonons, or boundaries \cite{deJongMolenkamp1995,AndreevKivelsonSpivak2011,LucasFong2018,PoliniGeim2020}. In this regime, the shear viscosity governs momentum diffusion and leaves direct signatures in nonlocal transport \cite{TorreEtAl2015,BandurinEtAl2016,SulpizioEtAl2019}, Poiseuille flow \cite{KumarEtAl2017,SulpizioEtAl2019,KuEtAl2020} and current-vortex formation \cite{LevitovFalkovich2016,BandurinEtAl2016,GuerreroBecerraEtAl2019}.

Small-angle (away from the magic angle) twisted bilayer graphene (SA-TBG) 
provides an especially appealing setting for this physics because the weak interlayer hybridization combined with externally applied displacement fields~\cite{Bandurin_prl_2022} can lift the (mini-)valley degeneracy while preserving a simple low-energy Dirac description \cite{LopesDosSantosEtAl2007,BistritzerMacDonald2011,CaoEtAl2018Insulator}. Valley splitting then becomes an additional tuning parameter for viscous transport, distinct from carrier density and temperature. Because it redistributes carriers between the two Dirac cones, and can even drive one valley across charge neutrality, it naturally changes the balance between intraband and electron-hole scattering channels and therefore need not affect the viscosity monotonically.

Recent advances in nonlocal transport measurements have begun to probe related valley-dependent thermodynamic and transport effects \cite{GorbachevEtAl2014,SuiEtAl2015,ChenEtAl2020,LiuEtAl2022}. Consequently, understanding how tunable valley polarization influences macroscopic fluid properties is highly timely. In particular, viscosity changes induced by valley splitting should manifest directly in tunable nonlocal resistance profiles, in the flow pattern of the current, and in the critical thresholds for current-vortex formation \cite{LevitovFalkovich2016, Bandurin_prl_2022, KumarEtAl2017, AharonSteinberg2022,SulpizioEtAl2019}.

\begin{figure}[t]
    \centering
    \includegraphics[width=0.9\linewidth]{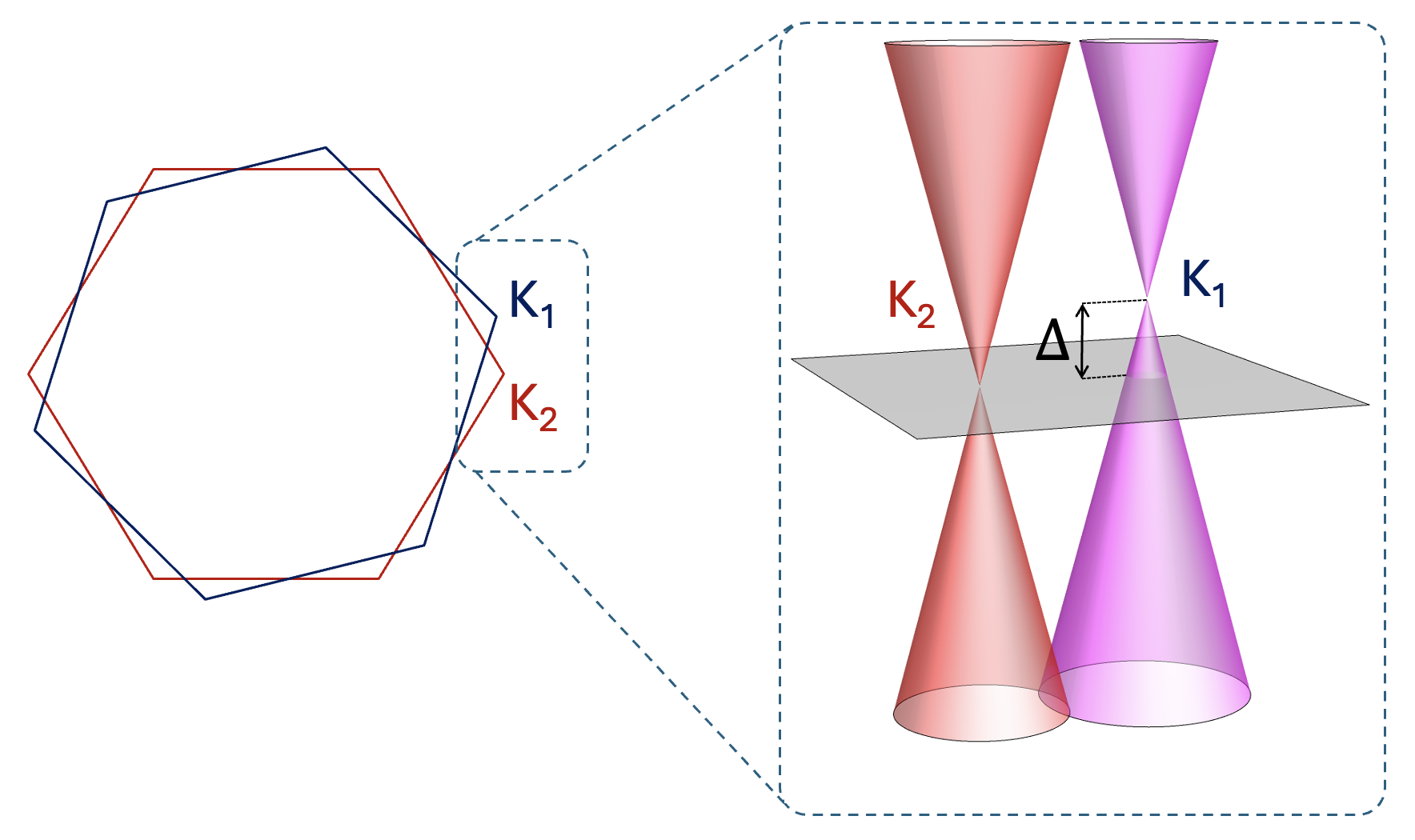}
    \caption{Schematic realization of the valley-polarized two-cone problem studied in this work. The two low-energy Dirac cones are offset by an energy splitting $\Delta$. This models the weak valley-dependent energy shift expected when the layers are weakly hybridized at small twist angle 
    {and a displacement field is applied}. The grey plane marks zero energy {in our calculation}.}
    \label{fig:twisted_cones}
\end{figure}

In this paper we use a linearized semiclassical Boltzmann approach to study how electron-electron scattering controls the shear and kinematic viscosities in this {valley-imbalanced} setting. The main result is that the viscosity of the two-cone Dirac system is a strongly nonmonotonic function of the valley splitting $\Delta$. As $\Delta$ is increased, the viscosity first rises, then develops a maximum and a second minimum, before increasing again in the strongly degenerate regime. 

We also study the role of carrier density and thermal effects. We explicitly focus on momentum-conserving electron-electron interaction. We assume, as it is appropriate in the hydrodynamic regime of graphene-based systems~\cite{BandurinEtAl2016,LucasFong2018,PoliniGeim2020}, that Umklapp processes, acoustic phonons, and static disorder play a subdominant role~\cite{HwangDasSarma2008,EfetovKim2010,DavisWuDasSarma2023}. We can then isolate the intrinsic viscous response driven purely by Coulomb interactions.

We also study monolayer graphene (MLG) and a two-dimensional electron gas (2DEG) as benchmarks and control cases. They serve two purposes: to test the kinetic framework against simpler limits and to isolate which features are specific to the SA-TBG problem. In particular, the MLG benchmark shows both the known features of electron viscosity and a subtle screening effect in the shear channel. Meanwhile, the parabolic-band model provides the conventional Fermi-liquid control case.

In MLG and the 2DEG, we also discuss the behavior of the shear and kinematic viscosities, $\eta$ and $\nu$, with particular emphasis on the viscosity minimum. In simple fluids, $\eta$ often exhibits a minimum at the crossover between liquid-like and gas-like behavior. When the mass density varies only weakly, $\nu$ does as well. Here we show that the shear viscosity of both MLG and the 2DEG likewise exhibits a minimum at the crossover between degenerate Fermi-liquid and nondegenerate Fermi-gas regimes. In MLG, however, the strong temperature dependence of the inertial mass density prevents a corresponding minimum in $\nu$, which instead decreases monotonically with temperature. This unusual behavior is a distinctive consequence of graphene's linear band structure and semimetallic character.

The paper is organized as follows. In Sec.~II we introduce the general kinetic framework for the calculation of shear viscosity. Section~III presents the valley-imbalanced two-cone Dirac system, its density constraint, the numerical viscosity curves, and analytic estimates for the extrema. Section~IV collects the benchmark systems: monolayer graphene and the parabolic-band 2DEG. Section~V contains conclusions and discussions. Technical derivations are collected in the Appendices.

\begin{figure}[t]
    \centering
    \includegraphics[width=0.7\linewidth]{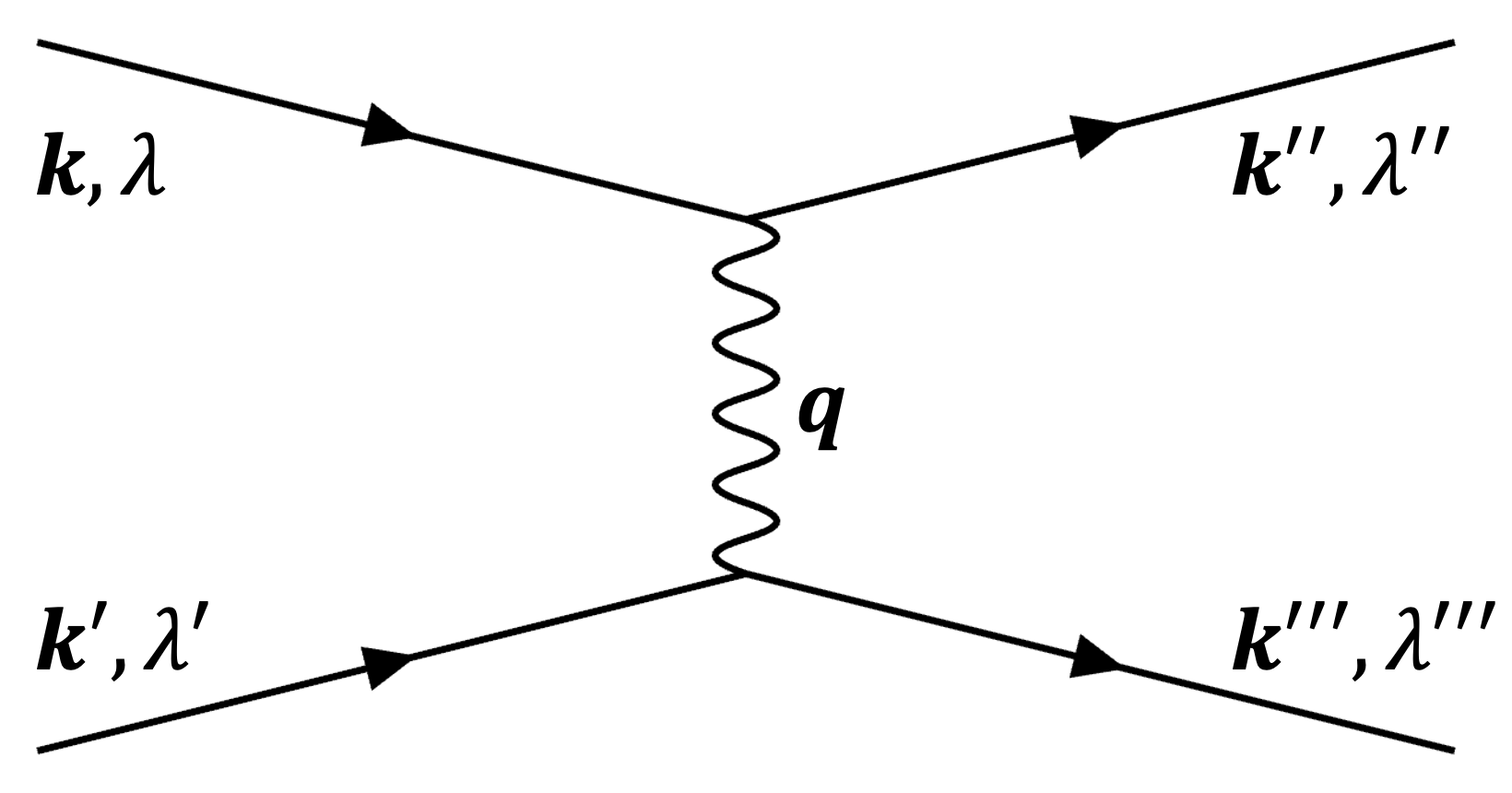}
    \caption{Representative two-body scattering process entering the collision integral. The four quasiparticle states are labelled by their momenta and band/valley indices. The wiggly line represents the (dynamically screened) Coulomb interaction, with ${\bm q}$ the transferred momentum. 
    }
    \label{fig:scattering}
\end{figure}

\section{General kinetic framework}\label{sec:graphene viscosity}
In this section we introduce the kinetic framework used throughout the paper to compute the shear viscosity of the electron fluids considered below. The starting point is the semiclassical Boltzmann equation for the quasiparticle distribution function $f_{{\bm k},\lambda}({\bm r})$, where $\lambda$ denotes a generic composite quasiparticle label collecting all discrete internal degrees of freedom. In the model-specific sections this general label will be specialized further, but throughout Sec.~\ref{sec:graphene viscosity} it is kept abstract.
Since our focus is the hydrodynamic regime, we retain only electron-electron scattering in the collision integral, which controls momentum-conserving relaxation of a nonuniform flow.
Within this approximation,
\begin{eqnarray} \label{eq:boltzmann_def}
&&{\bm v}_{{\bm k},\lambda}\cdot {\bm \nabla}_{\bm r} f_{{\bm k},\lambda}({\bm r}) -e
{\bm E}
\cdot\nabla_{{\bm k}} f_{{\bm k},\lambda}({\bm r}) ={\cal I}_{\rm ee}[f_{{\bm k},\lambda}({\bm r})]
~,
\end{eqnarray}
where ${\bm v}_{{\bm k},\lambda} = \nabla_{\bm k} \varepsilon_{{\bm k},\lambda}$. Here, $\varepsilon_{{\bm k},\lambda}$ is the band energy, while the collision integral is
\begin{widetext}
\begin{eqnarray} \label{eq:I_ee_def}
{\cal I}_{\rm ee}[f_{{\bm k},\lambda}({\bm r})] &=& \sum_{{\bm q},{\bm k}'} \sum_{\lambda',\lambda'',\lambda'''} W_{\lambda\lambda'\lambda''\lambda'''}({\bm k},{\bm k}', {\bm q}) \delta(\varepsilon_{{\bm k},\lambda}+\varepsilon_{{\bm k}',\lambda'}-\varepsilon_{{\bm k}+{\bm q},\lambda''}-\varepsilon_{{\bm k}'-{\bm q},\lambda'''})
\nonumber\\
&\times&
\big[ f_{{\bm k},\lambda} f_{{\bm k}',\lambda'} (1-f_{{\bm k}+{\bm q},\lambda''})(1-f_{{\bm k}'-{\bm q},\lambda'''}) - (1-f_{{\bm k},\lambda})(1-f_{{\bm k}',\lambda'}) f_{{\bm k}+{\bm q},\lambda''} f_{{\bm k}'-{\bm q},\lambda'''} \big]
~.
\end{eqnarray}
\end{widetext}
The scattering kernel $W_{\lambda\lambda'\lambda''\lambda'''}$ contains the dynamically screened Coulomb interaction $V_{\rm ee}({\bm q},\omega)=v_{\bm q}/\epsilon({\bm q},\omega)$, with $v_{\bm q}=2\pi e^2/(\kappa q)$ the bare Coulomb interaction and $\epsilon({\bm q},\omega)=1-v_{\bm q}\chi_{nn}({\bm q},\omega)$ the dielectric function. Here $\kappa$ is the dielectric constant of the surrounding medium and $\chi_{nn}$ is the density-density response function. The explicit formulas for $W_{\lambda\lambda'\lambda''\lambda'''}$ and the overlap factors are collected in App.~\ref{app:collintder}. Although Eq.~\eqref{eq:boltzmann_def} is written in a form that allows for external fields, all viscosity calculations below are performed at ${\bm E}=0$. For brevity, we set $\hbar=1$ in the intermediate kinetic derivations of Sec.~\ref{sec:graphene viscosity} and the Appendices, but restore factors of $\hbar$ in all final physical results and thermodynamic quantities.

Figure~\ref{fig:scattering} shows a representative momentum-conserving process contributing to the collision integral.

To calculate the shear viscosity, we consider a weakly nonuniform flow described by a local velocity field ${\bm u}({\bm r})$.
As a reference state we use a locally drifting distribution,
\begin{eqnarray}
f_{{\bm k},\lambda}({\bm r}) &=& n_{\rm F}\left(\frac{\varepsilon_{{\bm k},\lambda} - \mu - {\bm k}\cdot {\bm u}({\bm r})}{k_{\rm B} T}\right) +
\delta f_{{\bm k},\lambda}({\bm r})
\nn \\
&\equiv&
f^{(0)}_{{\bm k},\lambda}({\bm r}) + \delta f_{{\bm k},\lambda}({\bm r})
~,
\end{eqnarray}
where $n_{\rm F}(x) = (e^x+1)^{-1}$ is the Fermi distribution and $\delta f_{{\bm k},\lambda}({\bm r})$ is the nonequilibrium correction induced by spatial gradients of the flow.
Linearizing Eq.~\eqref{eq:boltzmann_def} in gradients of ${\bm u}({\bm r})$ and restricting to incompressible flow, $\partial_\rho u_\rho = 0$, we obtain
\be \label{eq:A}
\left(-\frac{\partial f^{(0)}(\varepsilon_{{\bm k},\lambda})}{\partial \varepsilon_{{\bm k},\lambda}}\right)
A^{\alpha\beta}_{{\bm k},\lambda}
\partial_\alpha u_\beta
=
{\cal I}_{\rm ee}^{({\rm lin})}[\delta f_{{\bm k},\lambda}({\bm r})],
\ee
where ${\cal I}_{\rm ee}^{({\rm lin})}$ is the linearized collision integral and {$A^{\alpha\beta}_{{\bm k},\lambda}~=~\varepsilon_{{\bm k},\lambda} (k^\alpha k^\beta/k^2 -\delta^{\alpha\beta}/2)$} is the traceless stress vertex that projects the kinetic equation onto the shear channel.
We approximate the nonequilibrium correction by the standard shear-mode Ansatz
\begin{eqnarray} \label{eq:deltaf_ansatz}
\delta f_{{\bm k},\lambda}({\bm r}) &=& \tau_v \left(-\frac{\partial f^{(0)}_{{\bm k},\lambda}}{\partial \varepsilon_{{\bm k},\lambda}}\right)
A^{\alpha\beta}_{{\bm k},\lambda} \partial_\alpha u_\beta,
\end{eqnarray}
where $\tau_v$ is the viscosity transport time.
In Eq.~(\ref{eq:deltaf_ansatz}) we approximate the perturbation to the equilibrium distribution by the leading shear harmonic and take $\tau_v$ to be independent of momentum and discrete index.

The dissipative stress tensor is proportional to the symmetrized traceless velocity gradient, $\Pi^{(\rm diss)}_{\alpha\beta}=2\eta u_{\alpha\beta}$.
Comparing this relation with the kinetic stress tensor obtained from the shear vertex and the correction $\delta f_{{\bm k},\lambda}$ gives
\be
\eta = {\cal D}\tau_v,
\qquad
\nu = \frac{\eta}{\rho_{\rm eff}}, \label{eq:viscpsities_def}
\ee
where ${\cal D}$ plays a role of viscosity Drude weight and $\rho_{\rm eff}$ is the effective inertial density appropriate to the model under consideration.

Substituting the Ansatz~\eqref{eq:deltaf_ansatz} into Eq.~\eqref{eq:I_ee_def} and projecting onto the shear channel yields a model-dependent expression for the collision rate $1/\tau_v$.
For the Dirac systems studied below, the final result is quoted in Sec.~\ref{sec:valley_polarized}, while the full reduction of the collision integral, including the explicit forms of the all the quantities introduced earlier, is given in App.~\ref{app:collintder}.

\section{Valley-polarized two-cone Dirac system}\label{sec:valley_polarized}

We now turn to an effective two-cone Dirac system in which an energy offset $\Delta$ separates the two low-energy cones. The setup is motivated by weakly hybridized SA-TBG, where (mini)valley-dependent energy shifts can be induced by the application of an interlayer potential difference ({\it i.e.} by a displacement field), while the low-energy spectrum still resembles two displaced Dirac cones. {In the explicit model below, the index $\Lambda=0,1$ labels these two displaced cones within a mini-Brillouin zone (mBZ). 
{Note, however, that the two valleys of the graphene Brillouin zone remain degenerate. Unless otherwise stated, all results are reported for a single valley in the Brillouin zone. The full valley multiplicity is obtained by multiplying by a factor of two. Throughout, the true valley index is suppressed, and the term ``valley'' is used as shorthand for ``mini-valley''.}
%
} Throughout this section we use the two-cone model as an effective description of the mini-valley-imbalanced viscous fluid introduced schematically in Fig.~\ref{fig:twisted_cones}.

\subsection{Model and density constraint}

{We denote the band index by $s=\pm 1$ and the displaced-valley index by $\Lambda=\{0,1\}$. When both labels need to be treated together, we group them into a composite index $a=(s,\Lambda)$ and write $\varepsilon_{{\bm k},a}\equiv\varepsilon_{{\bm k},s,\Lambda}$.}
The single-particle energy is taken to be
\begin{equation}
\varepsilon_{{\bm k},s,\Lambda}=s \hbar v_{\rm F} k+\Lambda\Delta ,
\end{equation}
We allow scattering between electrons belonging to the same valley
as well as to different ones. However, we assume that each electron preserves its valley index during the scattering process. That is, no intervalley scattering occurs, and the population of each valley is conserved.
This is appropriate when disorder is smooth on the lattice scale and the interaction is sufficiently long ranged, so that the large momentum transfer needed to switch valleys is suppressed. The corresponding collision-integral formulas are obtained by extending the framework of Sec.~II and are given in App.~\ref{app:valleys}.

{In equilibrium, the chemical potential $\mu$ is common to both displaced valleys, so the energy offset is accounted for by the effective chemical potential of each valley}
\begin{equation}
  \mu_\Lambda \equiv \mu-\Lambda\,\Delta ,
  \label{eq:muLambda_def}
\end{equation}
so that
$f^{(0)}(\varepsilon_{{\bm k},s,\Lambda}-\mu)=f^{(0)}(s\hbar v_{\rm F} k-\mu_\Lambda)$. Here $v_{\rm F}=10^6$~m/s is the Fermi velocity.
{Because the valley index is kept explicit throughout this section, the remaining internal degeneracy factor is the spin degeneracy $N = g_s=2$.}

For a two-dimensional Dirac fluid with internal degeneracy $N$, the net carrier density is
\begin{equation}
\begin{aligned}
    & n_{\rm D}(T,\mu)
    = \int_{0}^{\infty} d\varepsilon \, \frac{N\varepsilon}{2 \pi (\hbar v_{\rm F})^2}
    \left[
    n_{\rm F}\!\left(\frac{\varepsilon-\mu}{k_{\rm B}T}\right)-n_{\rm F}\!\left(\frac{\varepsilon+\mu}{k_{\rm B}T}\right)
    \right] \\
    &= \frac{N}{2 \pi} \left( \frac{k_{\rm B} T}{\hbar v_{\rm F}} \right)^2
    \left[ {\rm Li}_2\!\left( -e^{-\mu/(k_{\rm B} T)} \right) - {\rm Li}_2\!\left( -e^{\mu/(k_{\rm B} T)} \right) \right] ,
\end{aligned}
\label{eq:Dirac_density}
\end{equation}
while the corresponding energy density is
\begin{equation}
\begin{aligned}
    & \varepsilon_{\rm D}(T,\mu)
    = \int_0^\infty d\varepsilon \, \frac{N\varepsilon^2}{2 \pi (\hbar v_{\rm F})^2}
    \left[
    n_{\rm F}\!\left(\frac{\varepsilon-\mu}{k_{\rm B}T}\right)+n_{\rm F}\!\left(\frac{\varepsilon+\mu}{k_{\rm B}T}\right)
    \right] \\
    &= -\frac{N(k_{\rm B}T)^3}{\pi (\hbar v_{\rm F})^2}
    \left[ {\rm Li}_3\!\left( -e^{-\mu/(k_{\rm B} T)} \right) + {\rm Li}_3\!\left( -e^{\mu/(k_{\rm B} T)} \right) \right] .
\end{aligned}
\label{eq:Dirac_energy_density}
\end{equation}
For a scale-invariant Dirac fluid, the effective inertial density entering $\nu$ is naturally related to the enthalpy density $w_{\rm D}$ \cite{MullerSchmalianFritz2009,LucasFong2018},
\begin{equation}
    \rho_{\rm D}(T,\mu)
    = \frac{w_{\rm D}(T,\mu)}{v_{\rm F}^2}
    = \frac{d+1}{d}\frac{\varepsilon_{\rm D}(T,\mu)}{v_{\rm F}^2},
\label{eq:Dirac_rho}
\end{equation}
where $d=2$ is the number of spatial dimensions.

For later use, we define the single-valley carrier and inertial-mass densities by
\begin{equation}
  n_v(\mu,T)\equiv n_{\rm D}(T,\mu)\big|_{N=2},
  \quad
  \rho_v(\mu,T)\equiv \rho_{\rm D}(T,\mu)\big|_{N=2}.
  \label{eq:single_valley_def}
\end{equation}
The fixed gate-controlled density then becomes
\begin{equation}
  n_0
  =n_v(\mu_0,T)+n_v(\mu_1,T)
  =n_v(\mu,T)+n_v(\mu-\Delta,T),
  \label{eq:two_valley_density_constraint}
\end{equation}
which implicitly defines the global chemical potential $\mu(T,\Delta)$ for a given set of parameters $(n_0,T,\Delta)$.
The corresponding two-valley inertial density is
\begin{equation}
  \rho_{\rm eff}(T,\Delta)=\rho_v(\mu,T)+\rho_v(\mu-\Delta,T),
  \label{eq:two_cone_rho}
\end{equation}
and all kinematic and shear viscosities shown below are computed from Eq.~\eqref{eq:viscpsities_def}.

The corresponding viscosity Drude weight is
\be \label{eq:D_def}
    {\cal D} = N
    \sum_{a} \int \frac{d k}{2\pi} \frac{v_{\rm F}^2 k^3}{16 k_{\rm B} T \cosh^2\left(\frac{\varepsilon_{{\bm k}, a} -\mu}{2 k_{\rm B} T} \right)} ,
\ee
{where the composite index $a=(s,\Lambda)$ collects the band and valley labels.} 

The inverse viscosity transport time then assumes the following form
\begin{widetext}
\be  \label{eq:viscosity_time_final}
\frac{1}{\tau_v}
&=&
-\frac{2\pi v_{\rm F}^2}{8 {\cal D}} \sum_{a,b,c,d}
\int_{-\infty}^\infty d\omega
\int_{}^{} dq
\int_{}^{} dk
\int_{}^{} dk'
\frac{q}{2\pi} \frac{|V_{\rm ee}(q,\omega)|^2}{4 k_{\rm B} T \sinh^2(\frac{\omega}{2k_{\rm B} T})}
\frac{k}{2\pi}
\frac{k'}{2\pi}
{\cal I}_{a}(k,\omega,q)
{\cal I}'_{b}(k',\omega,q)
\nonumber\\
&\times&
[f^{(0)}(\varepsilon_{{\bm k},a}) - f^{(0)}(\varepsilon_{{\bm k},a}+\omega)]
[f^{(0)}(\varepsilon_{{\bm k}',b}) - f^{(0)}(\varepsilon_{{\bm k}',b} - \omega)]
{\cal M}_{ab}(k,k',\omega,q)
,
\ee
\end{widetext}
where ${\cal I}_{a}(k,\omega,q)$ and ${\cal I}'_{b}(k',\omega,q)$ are the angular phase-space factors generated by the two energy-conserving $\delta$-function integrals, while ${\cal M}_{ab}(k,k',\omega,q)$ is the corresponding shear-mode matrix element in the shear channel. The explicit expressions for ${\cal I}_{a}$, ${\cal I}'_{b}$, and ${\cal M}_{ab}$, together with the full reduction of the collision integral, are given in App.~\ref{app:collintder}. The corresponding valley-imbalance extension used in this section is summarized in App.~\ref{app:valleys}.

\subsection{Viscosity versus valley splitting}

For the numerical results presented below we fix the total charge carrier density to
$n_0=10^{11}$~cm$^{-2}$, unless explicitly stated otherwise. This value is well within the linear-band approximation~\cite{Bandurin_prl_2022}.


\begin{figure}[t]
    \centering
    \includegraphics[width=0.95\linewidth]{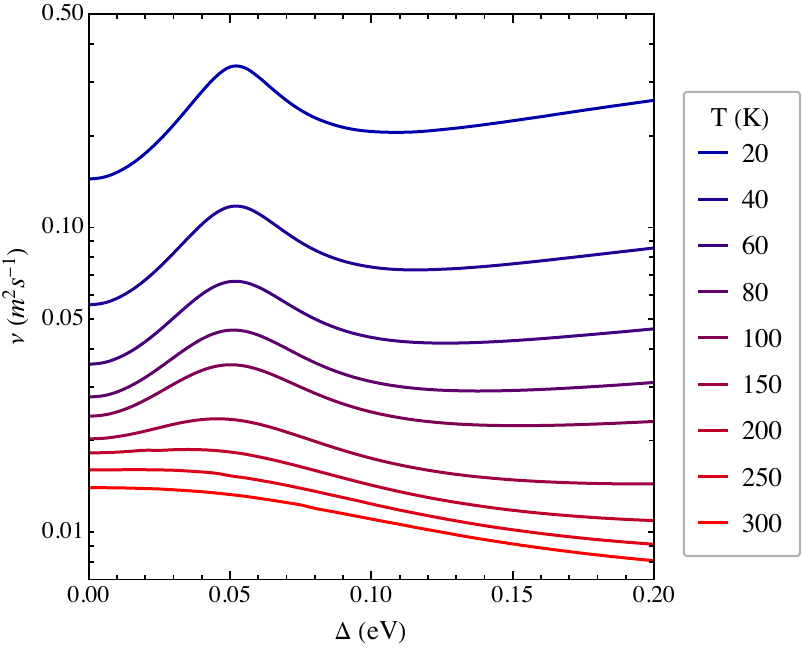}
    \caption{{Kinematic viscosity of the valley-imbalanced SA-TBG two-cone Dirac model as a function of the energy offset $\Delta$, at $n_0=10^{11}$~cm$^{-2}$ and for various temperatures. At low and intermediate temperatures the finite-$\Delta$ maximum and the local minimum are both clearly resolved. For $T>T_c$, the global maximum remains at $\Delta=0$, but the profile around the origin becomes broad, which makes the numerical extraction of a distinct finite-$\Delta$ peak increasingly noisy.}}
    \label{fig:kinematic_viscosity_second_minimum}
\end{figure}

The first qualitative effect of valley imbalance is seen in both the shear and kinematic viscosity.
For the sake of clarity, we present results for the shear viscosity in App.~\ref{app:deltamin}, Fig.~\ref{fig:viscosity_second_minimum}, while we focus here on the kinematic viscosity, shown in Fig.~\ref{fig:kinematic_viscosity_second_minimum}. 

\begin{figure}[t]
    \begin{tabular}{c}
    \begin{overpic}[width=0.95\linewidth]{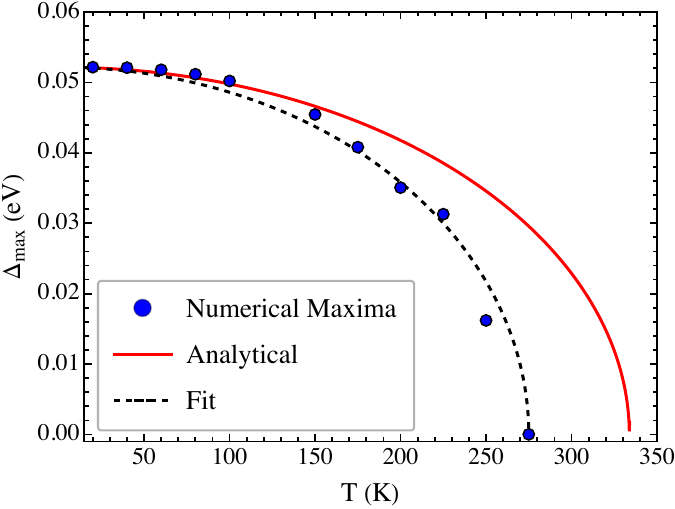}
    \put(1,157){{\large (a)}}
    \end{overpic}
    \\
    \begin{overpic}[width=0.95\linewidth]{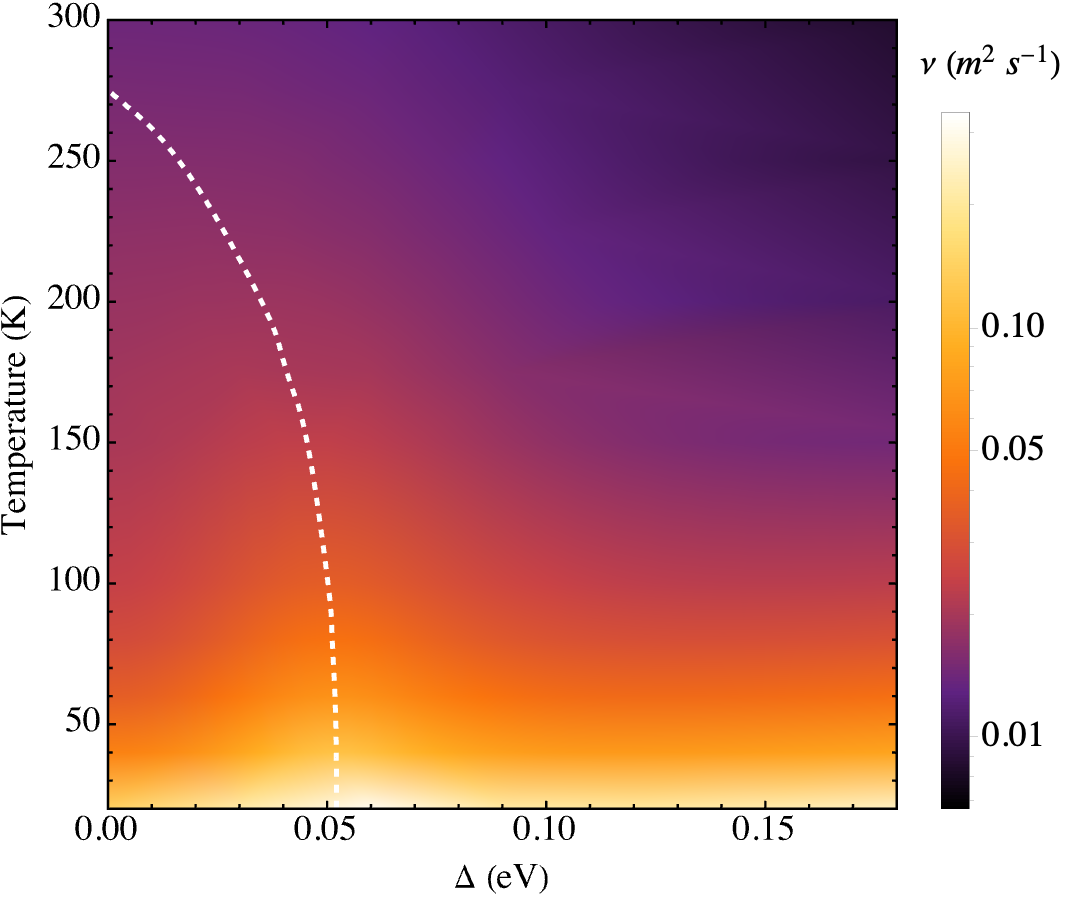}
    \put(1,175){{\large (b)}}
    \end{overpic}
    \end{tabular}
    \caption{\label{fig:delta_max_min}(a) Temperature dependence of the maximum of the kinematic viscosity, $\Delta_{\max}(T)=\arg\max_{\Delta}[\nu(\Delta,T)]$. The blue dots are the numerically extracted maxima, the red solid line is the low-temperature analytic estimate of Eq.~\eqref{eq:Deltamax_sommerfeld}, and the black dashed line is a phenomenological fit of the form ${\Delta^{\star}\sqrt{1-(T/T_c)^2}}$, where $T_c$ denotes the temperature at which the finite-$\Delta$ maximum collapses onto the broad maximum at $\Delta=0$. (b) {Density plot of $\nu(\Delta,T)$ in the $(\Delta,T)$ plane at fixed $n_0=10^{11}$~cm$^{-2}$. The dashed white line follows the extracted peak line and highlights how the finite-$\Delta$ maximum continuously approaches $\Delta=0$ as $T\to T_c$.} }
\end{figure}

The absolute minimum of the shear viscosity always remains at the unbalanced point $\Delta=0$.
{
However, in the temperature window $T<100$~K the viscosity develops a second minimum at finite $\Delta$.

For example, at $T=40$~K (Figs.~\ref{fig:kinematic_viscosity_second_minimum} and~\ref{fig:viscosity_second_minimum}) the shear viscosity first increases with valley splitting, reaches a maximum around $\Delta\approx 50$~meV, and then decreases towards a secondary minimum near $\Delta\approx 80$~meV. 
}
The {numerically-determined} maximum as a function of temperature $T$, $\Delta_{\max}(T)=\arg\max_\Delta[\nu(\Delta,T)]$, is shown in Fig.~\ref{fig:delta_max_min}. 
The behaviour of the secondary minimum is presented in Fig.~\ref{fig:deltamin_appendix}. Due to the weak dependence of the kinematic viscosity on $\Delta$, determining the secondary minimum at high temperatures is numerically challenging.

{Panel~(b) of Fig.~\ref{fig:delta_max_min} displays a density plot of the kinematic viscosity $\nu(\Delta,T)$ as a function of both valley-imbalance and temperature. This figure shows that the viscosity maximum (dashed line) tends toward $\Delta=0$ as the temperature increases.} The numerical data reveal a robust nonmonotonic dependence on valley splitting: the viscosity first rises as one valleys is depleted, then drops once the shifted valleys becomes hole-like and additional electron-hole scattering channels open, and finally increases again at larger $\Delta$ when both valleys are deep in the degenerate regime and Pauli blocking suppresses scattering.

\subsection{Physical interpretation}

For temperatures below $T_c \approx260$~K the curve $\nu(\Delta)$ has a clear maximum away from $\Delta=0$.
As temperature increases, the maximum of $\nu(\Delta,T)$ broadens and moves toward $\Delta=0$, so that its location becomes increasingly difficult to determine numerically.
Within our resolution, $\Delta_{\max}=0$ for all $T>T_c$.
Above this temperature we do not resolve a robust maximum at finite $\Delta$.
This behavior is consistent with a simple physical picture. For small $\Delta$, both valleys are electron doped and contribute comparable phase space for shear relaxation. Near $\Delta_{\max}$, the shifted valley is driven close to neutrality while the other valley remains electron doped; this suppresses the net contribution of the shifted cone and produces a peak in $\nu$. For $\Delta>\Delta_{\max}$, the shifted valley becomes hole doped and electron-hole scattering channels open, lowering the viscosity toward a second minimum at $\Delta_{\min}$. At still larger splitting, both valleys move deeper into the degenerate regime, phase-space restrictions and Pauli blocking dominate, and $\nu$ rises again.

The termination of the peak line at $T_c$ has the same origin. Once thermal broadening is already strong enough at $\Delta=0$ to wash out the distinction between a clearly electron-doped cone and a nearly neutral one, tuning $\Delta$ no longer creates a separate weak-relaxation regime. The finite-$\Delta$ maximum therefore broadens and disappears rather than persisting to arbitrarily high temperature.

\subsection{Analytic estimates for $\Delta_{\max}$ and $\Delta_{\min}$}

The numerical structure described above can be understood analytically in the degenerate regime. {For the single-cone density $n_v(\mu,T)$ defined in Eq.~\eqref{eq:single_valley_def}, the exact polylog expression \eqref{eq:Dirac_density} reduces, for $\mu\gg k_{\rm B}T$ ($\mu>0$), to the Sommerfeld expansion}
\begin{equation}
  n_v(\mu,T)\simeq
  \frac{g_s}{2\pi(\hbar v_{\rm F})^2}
  \left[
    \frac{\mu^2}{2}+\frac{\pi^2}{6}(k_{\rm B}T)^2
  \right],
  \label{eq:nD_sommerfeld}
\end{equation}
with corrections exponentially small in $-\mu/k_{\rm B}T$.
This approximation breaks down when $\mu \sim k_{\rm B}T$. A useful crossover marker is the intrinsic thermal density of electrons and holes at neutrality \cite{FritzEtAl2008,MullerFritzSachdev2008},
\begin{equation}
  n_{\rm th}(T)
  \equiv 
  \frac{g_s\pi}{6}\left(\frac{k_{\rm B}T}{\hbar v_{\rm F}}\right)^2,
  \label{eq:nth_neutrality}
\end{equation}
and the corresponding temperature scale $T^\ast$ defined by $n_{\rm th}(T^\ast)\approx n_0$.


\subsubsection{Evolution of the maximum $\Delta_{\max}(T)$.}

Numerically we find a pronounced maximum of $\nu(\Delta,T)$ at $\Delta=\Delta_{\max}(T)$.
Physically, this occurs when the shifted valley $\Lambda=1$ is tuned close to charge neutrality,
\begin{equation}
  \mu_1=\mu-\Delta \simeq 0
  \quad\Longrightarrow\quad
  \Delta_{\max}(T)\simeq \mu(T,\Delta_{\max}).
\label{eq:peak_condition}
\end{equation}
Under the condition $\mu_1=0$, the $\Lambda = 1$ valley carries no net charge, $n_v(0,T)=0$, and the density constraint \eqref{eq:two_valley_density_constraint} reduces to
\begin{equation}
  n_0 = n_v(\Delta_{\max},T).
\label{eq:peak_density_exact}
\end{equation}
Together with the exact polylog form \eqref{eq:Dirac_density}, Eq.~\eqref{eq:peak_density_exact} provides an implicit definition of $\Delta_{\max}(T)$.
In the degenerate regime $\Delta_{\max}\gg k_{\rm B}T$, using the Sommerfeld expansion \eqref{eq:nD_sommerfeld} yields
\begin{equation}
  n_0 \simeq \frac{g_s}{2\pi(\hbar v_{\rm F})^2}
  \left[
    \frac{\Delta_{\max}^2}{2}+\frac{\pi^2}{6}(k_{\rm B}T)^2
  \right].
\label{eq:peak_density_sommerfeld}
\end{equation}
Defining the fully polarized $T=0$ scale
\begin{equation}
  \Delta_0 \equiv \hbar v_{\rm F} \sqrt{\frac{4\pi n_0}{g_s}},
  \label{eq:Delta0_def}
\end{equation}
we obtain the compact estimate
\begin{equation}
  \Delta_{\max}(T)\simeq
  \sqrt{\Delta_0^2-\frac{\pi^2}{3}(k_{\rm B}T)^2},
  \qquad (\Delta_{\max}\gg k_{\rm B}T).
\label{eq:Deltamax_sommerfeld}
\end{equation}

Equation~\eqref{eq:Deltamax_sommerfeld} is therefore a degenerate-regime approximation. Once $T$ approaches $T^\ast$, or equivalently $\Delta_{\max}\sim k_{\rm B}T$, the full polylog relation \eqref{eq:Dirac_density} should be used instead. {Figure~\ref{fig:delta_max_min}(a) compares the numerically extracted maxima (dots) with the low-$T$ analytic result (red solid line) and with the phenomenological fit $\Delta^{\star}\sqrt{1-(T/T_c)^2}$ (black dashed line). For $T>T_c$ the maximum is in fact at $\Delta=0$, but the peak at the origin becomes very broad and the direct numerical extraction of a separate finite-$\Delta$ maximum becomes noisy; this is still visible in the full curves collected in Fig.~\ref{fig:kinematic_viscosity_second_minimum}, where the higher-$T$ profiles have their maxima at $\Delta=0$.} 


\subsubsection{Evolution of the local minimum $\Delta_{\min}(T)$.}

At larger valley splitting the shifted valley becomes hole doped, $\mu_1<0$, and strong electron-hole scattering channels become available. We empirically find a second minimum of $\nu(\Delta,T)$ at $\Delta=\Delta_{\min}(T)$. To obtain a simple analytic estimate, we assume that the minimum is reached once a moderately developed hole pocket has formed in the shifted valley. We therefore define the hole chemical potential
\begin{equation}
  \mu_h \equiv -\mu_1 = \Delta-\mu,
  \label{eq:muh_def}
\end{equation}
and parameterize the onset of the electron-hole regime as
\begin{equation}
  \mu_h \simeq \xi\,k_{\rm B}T,
  \label{eq:min_condition}
\end{equation}
where $\xi$ is a non-universal constant extracted from a fit to the numerical data. Writing $\mu_0\equiv\mu>0$ for the electron-doped valley and using Eq.~\eqref{eq:nD_sommerfeld} for the single-cone density, the degenerate electron-hole configuration gives
\begin{equation}
  n_0 \simeq \frac{g_s}{4\pi(\hbar v_{\rm F})^2}\left(\mu_0^2-\mu_h^2\right),
  \label{eq:eh_density_T2_cancellation}
\end{equation}
so that the $T^2$ corrections cancel in the net charge and the density constraint becomes algebraic:
\begin{equation}
  n_0=\frac{g_s}{4\pi(\hbar v_{\rm F})^2}\Big[\mu^2-(\Delta-\mu)^2\Big]
  =\frac{g_s}{4\pi(\hbar v_{\rm F})^2}\left(2\Delta\mu-\Delta^2\right).
\label{eq:eh_density_algebra}
\end{equation}
Solving for $\mu_h$ in terms of $\Delta$ and imposing the minimum condition \eqref{eq:min_condition} gives
\begin{equation}
  \frac{\Delta_{\min}^2-\Delta_0^2}{2\Delta_{\min}}=\xi k_{\rm B}T,
  \label{eq:Deltamin_equation}
\end{equation}
with solution
\begin{equation}
  \Delta_{\min}(T)\simeq
  \xi k_{\rm B}T+\sqrt{(\xi k_{\rm B}T)^2+\Delta_0^2}.
\label{eq:Deltamin_result}
\end{equation}
Equation~\eqref{eq:Deltamin_result} is expected to hold only when both {valleys within the mini-BZ} remain degenerate $|\mu|,|\mu-\Delta|\gg k_{\rm B}T$. {The estimation of the minima is susceptible to numerical errors because $\nu$ changes very slowly around $\Delta_{\rm min}$. Therefore, we report our findings on this in App.~\ref{app:deltamin}. }

\section{Monolayer graphene, parabolic 2DEG, and the role of screening}

To demonstrate that the nonmonotonic viscosity of the valley-imbalanced system is a unique consequence of intervalley carrier redistribution, we contrast our findings with the standard single-cone and parabolic-band limits. In this section, we analyze both 
MLG
and a conventional 
2DEG.
We use these model systems to also isolate 
the role of the temperature-dependent inertial density specific to Dirac fluids, and the critical necessity of static screening in the shear channel.

\subsection{Monolayer Graphene: Monotonicity and Screening}

In MLG, the low-energy spectrum is described by valley-degenerate massless Dirac fermions,
\begin{equation}
\varepsilon_{{\bm k},s}=s \hbar v_{\rm F} k ,
\end{equation}
where $s=\pm 1$ is the band index and $v_{\rm F}$ is the same Dirac velocity introduced in Sec.~\ref{sec:valley_polarized}. The total spin-valley degeneracy is $N=4$. Without the tunable valley splitting $\Delta$, the kinematic viscosity remains strictly monotonic with temperature. 

Furthermore, extending our kinetic framework to the MLG shear channel reveals a subtle but essential physical feature. While projecting the collision integral onto the shear mode successfully removes the collinear divergence associated with Dirac chirality, the bare Coulomb interaction still leaves a residual logarithmic divergence. As we show below, introducing static screening is not a secondary quantitative correction, but a strict physical requirement to restore the correct low-temperature Fermi-liquid-like asymptotics in the shear transport time.

\subsubsection{Neutrality-point Dirac fluid}

At charge neutrality, $\mu=0$, the viscosity Drude weight of Eq.~\eqref{eq:D_def} can be evaluated analytically (restoring all the units)
\be \label{eq:D_results_undoped}
{\cal D} \to \frac{9 \zeta(3)}{4} \frac{(k_{\rm B} T)^3}{2\pi (\hbar v_{\rm F})^2}.
\ee
Thus, in the Dirac fluid regime the prefactor entering the shear viscosity scales as $T^3$.
The corresponding viscosity transport time follows from Eq.~\eqref{eq:viscosity_time_final}. At neutrality, the only energy scale is set by temperature, and the collision integral can therefore be written in dimensionless form.
The result is
\be
\frac{1}{\tau_v}
=
\frac{k_{\rm B} T}{\hbar} C_v
,
\ee
where $C_v$ is a dimensionless constant obtained by evaluating the integral defined in Eq.~\eqref{eq:Cv_def}.
This linear-in-$T$ scaling is the basic result for the neutral Dirac fluid.
For a dielectric constant $\kappa=4$ and unscreened Coulomb interaction, $\epsilon(q,\omega)=1$, numerical evaluation of Eq.~\eqref{eq:Cv_def} gives $C_v=0.5864$,
which corresponds to $\tau_v=4.34\times 10^{-2}~{\rm ps}$
at room temperature, $T=300~{\rm K}$.
This value is consistent with previous calculations of the viscosity of undoped graphene~\cite{MullerSchmalianFritz2009,Narozhny2019}, where the linearized Boltzmann equation is expanded in two modes rather than in the single shear mode used here.
Converting the viscosity reported in Refs.~\onlinecite{MullerSchmalianFritz2009,Narozhny2019} into a transport time using Eq.~\eqref{eq:D_results_undoped}, one obtains $\tau_v=2.63\times 10^{-2}~{\rm ps}$,
namely a value smaller by a factor of about $1.5$, as expected from the inclusion of additional modes.


\begin{figure}[t]
    \begin{tabular}{c}
    \begin{overpic}[width=0.95\linewidth]{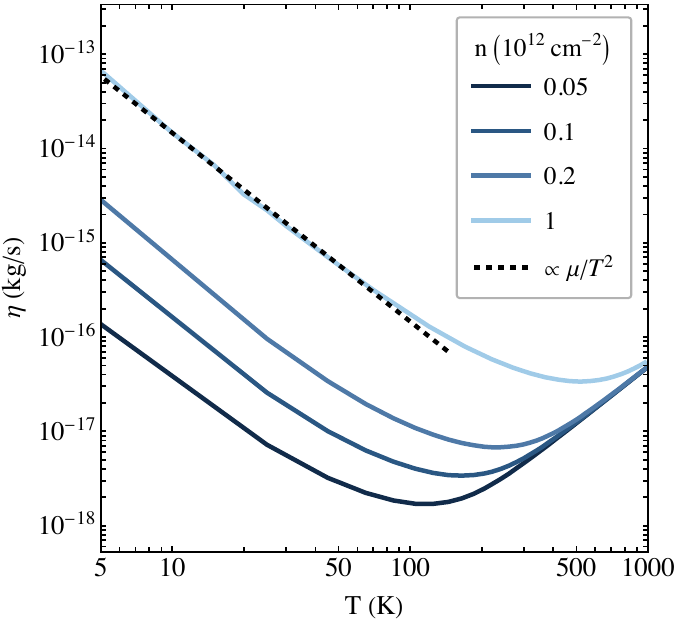}
    \put(1,180){{\large (a)}}
    \end{overpic}
    \\
    \begin{overpic}[width=0.95\linewidth]{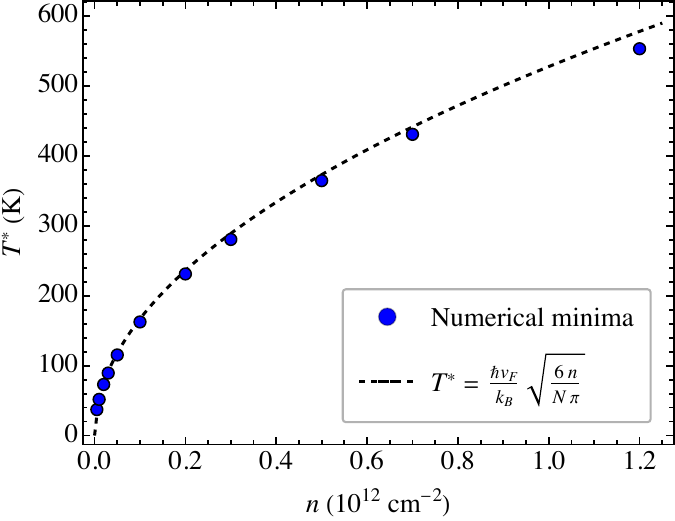}
\put(1,165){{\large (b)}}
\end{overpic}
\end{tabular}
    \caption{\label{fig:cazzo}(a) Shear viscosity of monolayer graphene as a function of temperature at fixed carrier densities. (b) Degenerate-to-Dirac crossover temperature $T^*(n)$ as function of carrier density in MLG. The dashed line follows Eq.~\eqref{eq:nth_neutrality}.}
\end{figure}

We can also consider the impact of statically screened Coulomb interactions. In the long-wavelength and zero-frequency limit, we approximate the dielectric function as
\begin{equation}
\epsilon(q,\omega)\to 1-v_{\bm q}\chi_{nn}^{(0)}(q\to 0,\omega=0;T),
\end{equation}
where
\begin{align}
    &{chi_{nn}^{(0)}(q\to 0,\omega=0;T) = -\frac{\partial n}{\partial \mu}}  \nonumber \\
    &{=  \frac{4 k_B T}{\pi (\hbar v_F)^2}\ln \left[ 2 \cosh \left( \frac{\mu}{2 k_B T} \right) \right]} \label{eq:screeining}
\end{align}
is the static density response of noninteracting monolayer graphene. At charge neutrality, $\mu=0$, Eq.~\eqref{eq:screeining} reduces to
\begin{equation}
    \chi_{nn}^{(0)}(\mu=0) = \frac{4 \ln(2) k_B T}{\pi (\hbar v_F)^2} .
\end{equation}
With the static screening, $C_v=0.0648$,
which gives $\tau_v=0.4~{\rm ps}$, in good agreement with available experimental estimates~\cite{BandurinEtAl2016,SulpizioEtAl2019,KuEtAl2020}.
We also compare our results for the neutral Dirac fluid with the Kovtun-Son-Starinets bound set by the holographic duality \cite{KovtunSonStarinets2005,MullerSchmalianFritz2009}. 
Using Eq.~\eqref{eq:D_results_undoped}, we obtain
\begin{equation}
    \eta/s \approx 2.7\, \frac{\hbar}{4\pi k_{\rm B}},
\end{equation}
where
\begin{equation}
s \approx \frac{3 \zeta(3) N k_{\rm B}^3 T^2}{2\pi (\hbar v_{\rm F})^2}
\end{equation}
is the entropy density of the noninteracting Dirac fluid.
This indicates that charge-neutral graphene remains an electronic fluid with remarkably low specific viscosity. Although it does not saturate the bound, it still deserves to be regarded as a nearly perfect quantum liquid~\cite{MullerSchmalianFritz2009}.

\begin{figure}[t]
    \begin{tabular}{c}
    \begin{overpic}[width=0.95\linewidth]{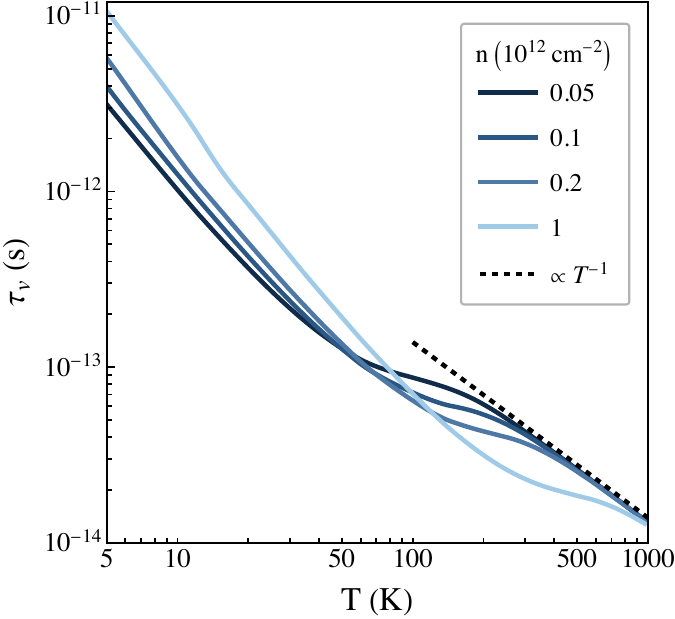}
    \put(1,180){{\large (a)}}
    \put(60,200){{\large Bare interaction}}
    \end{overpic}
    \\
    \begin{overpic}[width=0.95\linewidth]{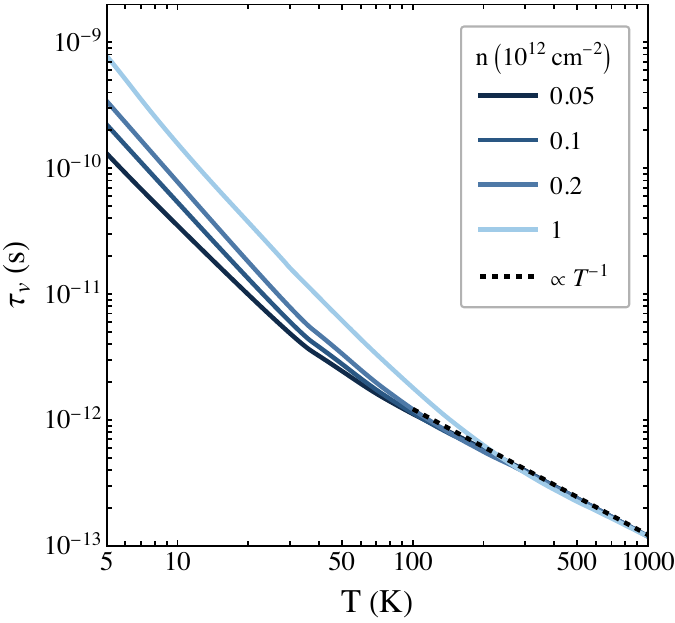}
\put(1,165){{\large (b)}}
\put(50,195){{\large Screened interaction}}
\end{overpic}
\end{tabular}
    \caption{\label{fig:nu01} Viscosity collision time, $\tau_v$, of monolayer graphene as a function of temperature at fixed carrier densities. Panel (a): bare Coulomb interaction; panel (b): screened interaction. Unlike the shear viscosity in Fig.~\ref{fig:cazzo}(a), both $\tau_v$ and the kinematic viscosity remain monotonic in the temperature range shown. The two differ only by the constant factor $v_{\rm F}^2/4$.}
\end{figure}


\subsubsection{Finite density and the minimal-viscosity regime}

We now turn to doped monolayer graphene.
At fixed carrier density, the chemical potential is temperature dependent, $\mu=\mu(n_0,T)$ and can be determined by inverting the general Dirac density formula,
\begin{equation}
    n_0 = n_{\rm D}(T,\mu)\big|_{N=4},
\end{equation}
with typical densities in the hydrodynamic regime in the range $n_0\sim 10^{11}$--$10^{12}$~cm$^{-2}$.
The resulting temperature dependence of the shear viscosity is shown in Fig.~\ref{fig:cazzo}(a) for a range of carrier densities.
We find a pronounced minimum at $T \approx 100~{\rm K}$ {for $n=10^{11}$ cm$^{-2}$, with its location shifting systematically with carrier density across the plotted range,} and $\eta_{\rm min}\sim 10^{-19}~{\rm kg/s}$, consistent with \cite{PrincipiEtAl2016,PongsanganganEtAl2024}.

As discussed previously in Sec.~\ref{sec:valley_polarized}, the crossover from Fermi-liquid to Dirac plasma happens when density of thermally excited carriers ($n_{\rm th}$ from Eq.~\eqref{eq:nth_neutrality}) becomes comparable to the density of carriers induced by electrostatic doping, $n_0$. We present the temperature of this crossover, $T^*$, as a function of doping, $n_0$, in Fig.~\ref{fig:cazzo}(b) defined as $T^* = \arg \min [\eta(T)]$. The dashed line is the solution to the equation 
\be
    n_{\rm th}(T) = n_0.
\ee

For monolayer graphene we use the Dirac inertial density with the full spin-valley degeneracy,
\begin{equation}
\rho_{\rm eff}^{\rm MLG}=\rho_{\rm D}(T,\mu)\big|_{N=4}. \label{eq:rho_mlg}
\end{equation}
The corresponding hydrodynamic mass per particle is then 
$M_{e,\text{HD}}=\rho_{\rm eff}^{\rm MLG}/n_0$.
In the fully degenerate limit, $T\to 0$, this reduces to the standard kinematic mass
\begin{equation}
m^*=\frac{\hbar k_F}{v_{\rm F}}=\frac{\hbar}{v_{\rm F}}\sqrt{\pi n_0}.
\end{equation}


The temperature dependence of the viscosity transport time (or, equivalently, of the kinematic viscosity after the constant conversion factor $v_{\rm F}^2/4$) is shown in Fig.~\ref{fig:nu01}. To illustrate more clearly the role of screening, we show both the bare-interaction result Fig.~\ref{fig:nu01}(a) and the statically screened result Fig.~\ref{fig:nu01}(b). In both cases, $\tau_v(T)$ and $\nu(T)$ remain monotonic over the temperature range considered here, in contrast to the shear viscosity. This difference originates from the temperature dependence of the effective inertial density in Eq.~\eqref{eq:rho_mlg} and is specific to the linear Dirac dispersion of graphene. 

In the unscreened case Fig.~\ref{fig:nu01}(a), one can see a broad shoulder at intermediate temperatures. This reflects the same competition that produces the minimum in $\eta(T)={\cal D}(T)\tau_v(T)$: the rapid decrease of $\tau_v(T)$ with temperature is partly compensated by the crossover of the viscosity Drude weight ${\cal D}(T)$. The competition is not strong enough to generate a true minimum in $\tau_v$ (and hence in $\nu$), but it does leave behind a plateau-like feature. Once static screening is included, this feature disappears and the curves evolve into a smoother monotonic decline. Screening therefore does more than regularize the residual logarithmic divergence of the bare Coulomb interaction: it also has a visible quantitative impact on the overall shape of the transport curves, removing features around the crossover that are still present in the unscreened calculation.

It is instructive to compare these values with those of conventional three-dimensional liquids.
The typical minimal kinematic viscosities in ordinary fluids lie in the range $10^{-7}$--$10^{-6}\,{\rm m}^2/{\rm s}$, whereas for graphene we obtain values of order $10^{-2}\,{\rm m}^2/{\rm s}$.
In this sense, the electron fluid in graphene is several orders of magnitude more viscous when measured in kinematic units.

\begin{figure}[t]
    \centering
    \includegraphics[width=0.95\linewidth]{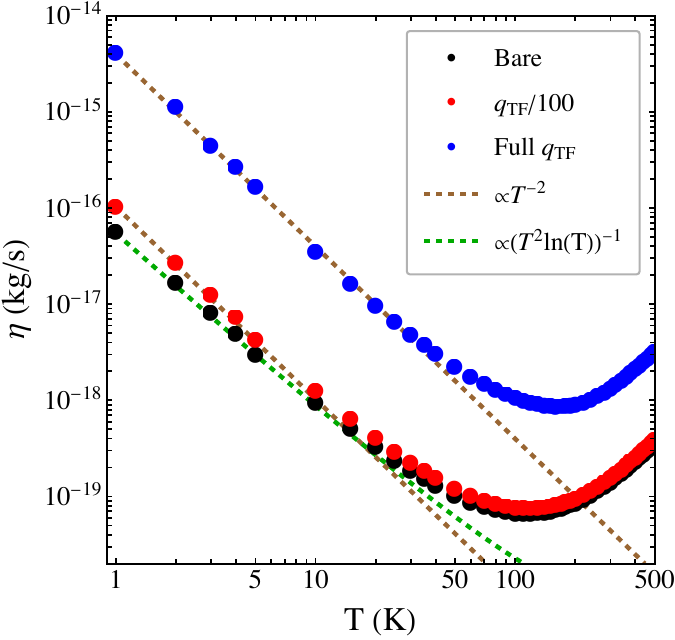}
    \caption{
    Temperature dependence of the shear viscosity in monolayer graphene for different screening mechanisms. The low-temperature asymptotics is highly sensitive to screening. A weak static screening is sufficient to recover the conventional $1/T^2$ asymptotics.
    }
    \label{fig:shear_screening_placeholder}
\end{figure}

\subsubsection{Screening and low-$T$ asymptotics in the shear channel}

The role of screening in the shear channel deserves to be stated separately. First, we introduce the Thomas-Fermi wavevector 
\begin{equation}
    q_{\rm TF} = 2 \pi \alpha \frac{\partial n}{\partial \mu}, 
\end{equation}
so the statically screened Coulomb interaction can be written as 
\begin{eqnarray}
    V_{\rm scr}(q) = \frac{2 \pi \alpha}{q + q_{\rm TF}}. \label{eq:screeningdef}
\end{eqnarray}
The shear mode removes the collinear divergence associated with the Dirac-wavefunction chirality factor, but this is not yet the full low-temperature story. Our calculation shows that with a bare Coulomb interaction a residual logarithmic divergence survives, so that the low-$T$ behavior remains quasiparticle-lifetime-like, $\sim 1/[T^2 \ln T]$ (Fig.~\ref{fig:shear_screening_placeholder}). Once the interaction is statically screened [Eq.~\eqref{eq:screeningdef}], the correct asymptotic behavior crosses over to the Fermi-liquid-like $T^{-2}$ form. In practice, even a very small fraction of the full Thomas-Fermi screening (red points in Fig.~\ref{fig:shear_screening_placeholder}) already drives the numerical result close to the screened asymptote.

\subsection{Parabolic-Band 2DEG: The Fermi-Liquid Contrast}

To isolate the features of the shear and kinematic viscosities that are unique to the Dirac dispersion, we contrast our results with a parabolic-band model, given by the Hamiltonian
\begin{equation}
    H = \frac{k^2}{2m},
\end{equation}
which is appropriate for a conventional 2DEG. Unlike the Dirac fluid, where the inertial density is intrinsically temperature-dependent, the 2DEG provides a standard Fermi-liquid baseline where the mass density is strictly fixed by the carrier concentration as $\rho = n m$. 

Furthermore, analyzing the 2DEG collision integral reveals a divergence at small momentum transfers for the bare Coulomb interaction. Consequently, static screening is an essential regularization step to obtain a finite transport time. The results show that, unlike in the Dirac systems, both kinematic and shear viscosities develop a minimum as functions of temperature. 

The kinetic calculation follows the same steps as in Sec.~II: one linearizes the Boltzmann equation in the shear channel, projects onto the corresponding tensor structure, and extracts the viscosity transport time from the linearized collision integral.
Since the algebra is analogous but technically lengthy, we defer the derivation to App.~\ref{app:2deg} and quote here the final result.
We denote quantities referring to the parabolic-band 2DEG by the subscript ``p'':
\begin{widetext}
\bea\label{eq:tau_parabolic}
    \frac{1}{\tau_{\rm p}} = \frac{-2\pi }{8 m^2 k_{\rm B} T {\cal D}_{\rm p}} && \int dq dk dk' \int  d\omega \frac{q}{2\pi} \frac{k}{2\pi} \frac{k'}{2\pi} \frac{|V_{\rm ee}({\bm q},\omega)|^2}{4 \sinh^2 \left( \frac{\omega}{2 k_{\rm B}T} \right)}[ f^{(0)}(\varepsilon_{\bm k}) - f^{(0)}(\varepsilon_{\bm k}+\omega) ] [ f^{(0)}(\varepsilon_{\bm k'}) - f^{(0)}(\varepsilon_{\bm k'} - \omega) ] \nonumber \\
    && \times {\cal I}_{\rm p}(k,q,\omega) {\cal I}_{\rm p}(k',q,-\omega) {\cal M}_{\rm p},
\eea
\end{widetext}
where
\be
    {\cal M}_{\rm p} = 2 ( k^2 + k'^2 ) q^2 - q^4 - 4 m^2 \omega^2,
\ee
and the corresponding Drude 
weight is
\be
    {\cal D}_{\rm p}(\mu, T) = - \frac{ m (k_{\rm B}T)^2}{4\pi}{\rm Li}_2 \left( -e^{\mu/k_{\rm B}T} \right),
\ee
The angular integral entering Eq.~\eqref{eq:tau_parabolic} is
\bea
    {\cal I}_{\rm p}(k,q,\omega)
    &=& \frac{2m}{\pi}\frac{\theta \left( k - \left| \frac{q^2-2m\omega}{2q} \right| \right)}{\sqrt{4k^2q^2- \left( q^2 - 2 m\omega \right) ^2 }}.
\eea

Equation~\eqref{eq:tau_parabolic} provides the parabolic-band analogue of the monolayer result.
The structure is similar. In the doped regime, the $q$ integral generated by the collision kernel is divergent at its lower bound for the bare Coulomb interaction.
For this reason, screening is not a secondary quantitative correction here, but an essential part of the calculation.
We use the same static screening as defined in Eq.~\eqref{eq:screeningdef}. This screened form controls the magnitude of the transport time and makes the low-temperature asymptotics well defined.

The low-temperature limit $\mu \gg k_{\rm B}T$ can be obtained analytically by restricting the momenta to the Fermi surface, as shown in App.~\ref{app:2deg_doped}. In this regime Eq.~\eqref{eq:tau_parabolic} reduces to
\be
    \frac{1}{\tau_{\rm p}} = \frac{2 \pi \alpha^2}{3}\frac{m}{\varepsilon_F} \frac{(k_{\rm B} T)^2}{\varepsilon_F} \int_{0}^{2k_F} \frac{q \, dq}{(q+q_{\rm TF})^2},
\ee
which makes the two main features of the parabolic-band problem explicit. First, the viscosity relaxation rate follows the conventional Fermi-liquid scaling $1/\tau_{\rm p} \propto T^2$ at low temperature. Second, the remaining momentum integral depends explicitly on the screening wavevector $q_{\rm TF}$, showing that forward scattering must be regularized at the level of the transport time.

\section{Conclusions}

The central result of this paper is that a valley-imbalanced two-cone Dirac fluid, motivated by weakly hybridized small-angle twisted graphene, exhibits a strongly nonmonotonic viscosity as a function of valley splitting $\Delta$. As the latter is increased, the kinematic viscosity first rises, then develops a local maximum and a second minimum, and finally increases again in the strongly degenerate regime. The characteristic lines $\Delta_{\max}(T)$ and $\Delta_{\min}(T)$ admit simple analytic estimates: {the maximum occurs when the shifted valley is driven close to charge neutrality, whereas the local minimum appears once that valley becomes hole doped and electron-hole scattering channels become effective.}

{
We also calculate the viscosity of monolayer graphene and a parabolic-band 2DEG. In monolayer graphene at finite density, the shear viscosity develops a minimum while the kinematic viscosity remains monotonic. This behavior is unusual and reflects the pronounced temperature dependence of the inertial mass density at fixed carrier density. By contrast, in many simple fluids, both shear and kinematic viscosity exhibit a minimum near the crossover between dense liquid-like and dilute gas-like regimes. In the liquid regime, the viscosity typically rises rapidly as the temperature is lowered and may approximately follow an activated form, while in the dilute-gas regime kinetic theory gives $\eta \propto T^{1/2}$ and, at fixed pressure, $\nu \propto T^{3/2}$~\cite{TrachenkoBrazil2015,LandauLifshitzFluid}.

In a 2DEG, by contrast, the mass density is fixed by the product of carrier density and effective mass, so the shear and kinematic viscosities have the same temperature dependence and both exhibit a minimum. Although this superficially resembles the behavior of simple fluids, the low-temperature viscosity of a Fermi liquid is qualitatively different: rather than increasing exponentially, it grows algebraically, $\eta \propto T^{-2}$. The underlying reason is the presence of a Fermi surface: Pauli blocking severely restricts the phase space for low-energy scattering and, correspondingly, renders generic quasiparticle interactions irrelevant in the renormalization-group sense~\cite{Shankar1994}. As a result, the electron-electron scattering rate vanishes algebraically as $T \to 0$, leading in turn to an algebraic divergence of the viscosity.
}

Finally, we emphasized a technical aspect related to screening. In monolayer graphene, the projection onto shear modes suppresses the collinear singularity that would otherwise arise in the Dirac collision integral, but the bare Coulomb interaction still produces a residual logarithmic divergence. Screening is therefore necessary to recover the correct low-temperature asymptotics. In the parabolic-band 2DEG, by contrast, the low-temperature transport is Fermi-liquid-like, and static screening is required to regularize the transport integral and obtain a finite viscosity relaxation time.

\begin{acknowledgments}
We acknowledge support from the Leverhulme Trust under the grant agreement RPG-2023-253.
\end{acknowledgments}

\begin{widetext}

\appendix

\section{Additional numerical results for the valley-imbalanced two-cone model}\label{app:deltamin}
{This appendix collects numerical information complementary to Sec.~\ref{sec:valley_polarized}. Figure~\ref{fig:viscosity_second_minimum} shows a representative shear-viscosity curve
Figure~\ref{fig:deltamin_appendix} summarizes the extracted local-minimum line $\Delta_{\min}(T)$ together with the guiding estimate of Eq.~\eqref{eq:Deltamin_result}. The local minimum is a robust qualitative feature, but its precise numerical location is less stable because $\nu(\Delta,T)$ varies only weakly near $\Delta_{\min}$.}

\begin{figure}[t]
    \centering
    \includegraphics[width=0.55\linewidth]{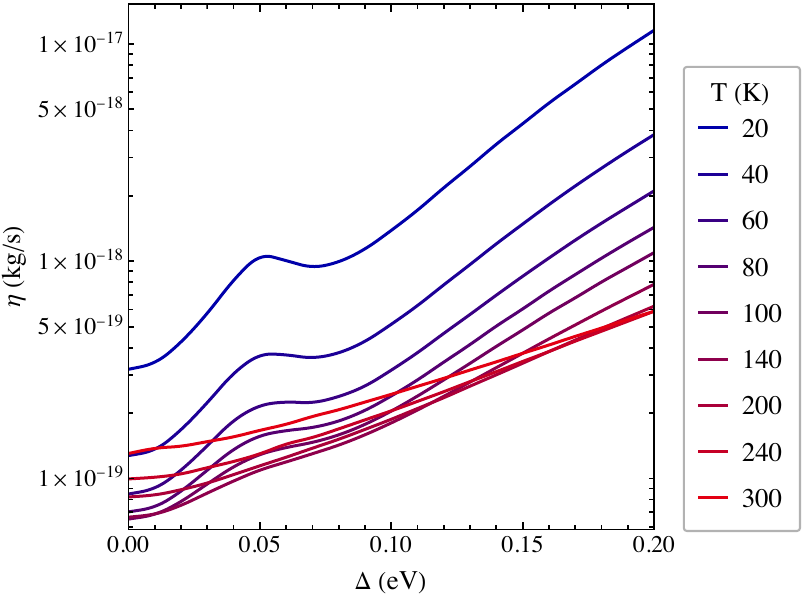}
    \caption{Representative shear-viscosity curves of the valley-imbalanced SA-TBG two-cone Dirac model at  $n_0=10^{11}$~cm$^{-2}$, as a function of the energy offset $\Delta$. The absolute minimum remains at $\Delta=0$, while a weaker secondary minimum develops at finite $\Delta$ in the same temperature window discussed in Sec.~\ref{sec:valley_polarized}.}
    \label{fig:viscosity_second_minimum}
\end{figure}

\begin{figure}[t]
    \centering
    \includegraphics[width=0.55\linewidth]{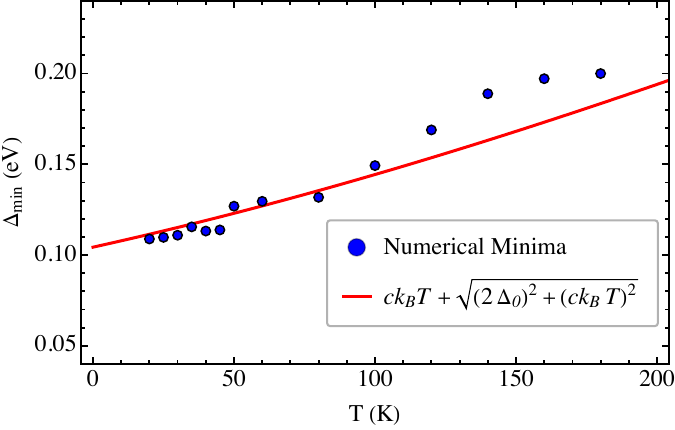}
    \caption{{Temperature dependence of the local-minimum line $\Delta_{\min}(T)$. Blue dots: numerically extracted local minima of $\nu(\Delta,T)$. Red solid line: guiding estimate from Eq.~\eqref{eq:Deltamin_result}. The overall upward trend is robust, but the precise position of the minimum is more susceptible to numerical noise than the corresponding peak line because the kinematic viscosity varies only weakly around $\Delta_{\min}$.}}
    \label{fig:deltamin_appendix}
\end{figure}

\clearpage

\section{Derivation of the collision integral}\label{app:collintder}
The quantity $A^{\alpha\beta}_{{\bm k},\lambda}$ of Eq.~\eqref{eq:A} is defined as follows
\be
A^{\alpha\beta}_{{\bm k},\lambda} &\equiv& \lambda v_{\rm F} k \left(\frac{k^\alpha k^\beta}{k^2} - \frac{1}{2}\delta^{\alpha\beta}\right)
=
\frac{\lambda v_{\rm F} k}{2} \big[ \cos(2\varphi_{\bm k}) \sigma_z^{\alpha\beta}  + \sin(2\varphi_{\bm k}) \sigma_x^{\alpha\beta} \big].
\ee
Note that $A^{\alpha\beta}_{{\bm k},\lambda}$ contains only transverse fluctuations. 

\subsection{Solving the linearized Boltzmann equation}
To find $\tau_v$ we multiply Eq.~\eqref{eq:A} by $A^{\mu\nu}_{{\bm k},\lambda}$ and sum over ${\bm k}$ and $\lambda$.
Eq.~\ref{eq:A} becomes
\be \label{eq:Boltzmann_lin_rhs_1}
{\cal D} u_{\mu\nu} &=& 
\frac{2\pi\tau}{4 k_{\rm B} T} \sum_{{\bm q},{\bm k},{\bm k}'} \sum_{\lambda,\lambda',\lambda'',\lambda'''} |V_{\rm ee}({\bm q},\varepsilon_{{\bm k},\lambda} - \varepsilon_{{\bm k}+{\bm q},\lambda''})|^2 
{\cal F}({{\bm k},\lambda}; {\bm k}+{\bm q},\lambda'') {\cal F}({{\bm k}',\lambda'}; {\bm k}'-{\bm q},\lambda''') 
\nonumber\\
&\times&
\delta(\varepsilon_{{\bm k},\lambda}+\varepsilon_{{\bm k}',\lambda'}-\varepsilon_{{\bm k}+{\bm q},\lambda''}-\varepsilon_{{\bm k}'-{\bm q},\lambda'''})
f_{{\bm k},\lambda}^{(0)} f_{{\bm k}',\lambda'}^{(0)} (1-f_{{\bm k}+{\bm q},\lambda''}^{(0)})(1-f_{{\bm k}'-{\bm q},\lambda'''}^{(0)})
\nonumber\\
&\times&
( A^{\mu\nu}_{{\bm k},\lambda} +  A^{\mu\nu}_{{\bm k}',\lambda'} -  A^{\mu\nu}_{{\bm k}+{\bm q},\lambda''} -  A^{\mu\nu}_{{\bm k}'-{\bm q},\lambda'''})( A^{\alpha\beta}_{{\bm k},\lambda} +  A^{\alpha\beta}_{{\bm k}',\lambda'} -  A^{\alpha\beta}_{{\bm k}+{\bm q},\lambda''} -  A^{\alpha\beta}_{{\bm k}'-{\bm q},\lambda'''}) \partial_\alpha u_\beta.
\ee
In Eq.~\ref{eq:Boltzmann_lin_rhs_1} we have symmetrized the linearized collision integral. We can further simplify the right-hand side of Eq.~\ref{eq:Boltzmann_lin_rhs_1} by noting that, upon integration, odd functions of $\varphi_{\bm q}$ vanish.
Therefore, we get
\be \label{eq:Boltzmann_lin_rhs_2}
{\cal D} u_{\mu\nu} 
&=& 
\frac{2\pi v_{\rm F}^2 \tau}{16 k_{\rm B} T} \sum_{{\bm q},{\bm k},{\bm k}'} \sum_{\lambda,\lambda',\lambda'',\lambda'''} |V_{\rm ee}({\bm q},\varepsilon_{{\bm k},\lambda} - \varepsilon_{{\bm k}+{\bm q},\lambda''})|^2 
{\cal F}({{\bm k},\lambda}; {\bm k}+{\bm q},\lambda'') {\cal F}({{\bm k}',\lambda'}; {\bm k}'-{\bm q},\lambda''') 
\nonumber\\
&\times&
\delta(\varepsilon_{{\bm k},\lambda}+\varepsilon_{{\bm k}',\lambda'}-\varepsilon_{{\bm k}+{\bm q},\lambda''}-\varepsilon_{{\bm k}'-{\bm q},\lambda'''})
f_{{\bm k},\lambda}^{(0)} f_{{\bm k}',\lambda'}^{(0)} (1-f_{{\bm k}+{\bm q},\lambda''}^{(0)})(1-f_{{\bm k}'-{\bm q},\lambda'''}^{(0)})
\nonumber\\
&\times&
\Big\{
\big[\lambda k \cos(2\varphi_{{\bm k}}) + \lambda' k' \cos(2\varphi_{{\bm k}'}) - \lambda'' |{\bm k}+{\bm q}|
\cos(2\varphi_{{\bm k}+{\bm q}}) - \lambda''' |{\bm k}' -{\bm q} |\cos(2\varphi_{{\bm k}'-{\bm q}}) \big]^2
\sigma_z^{\mu\nu} \sigma_z^{\alpha\beta}  
\nn \\
&+&
\big[\lambda k \sin(2\varphi_{{\bm k}}) + \lambda' k' \sin(2\varphi_{{\bm k}'}) - \lambda'' |{\bm k}+{\bm q}|
\sin(2\varphi_{{\bm k}+{\bm q}}) - \lambda''' |{\bm k}' -{\bm q} | \sin(2\varphi_{{\bm k}'-{\bm q}}) \big]^2
\sigma_x^{\mu\nu} \sigma_x^{\alpha\beta}  
\Big\}
\partial_\alpha u_\beta.
\ee
Shifting $\varphi_{{\bm k}} \to \pi/4 - \varphi_{{\bm k}}$, $\varphi_{{\bm k}'} \to \pi/4 - \varphi_{{\bm k}'}$ and $\varphi_{{\bm q}} \to \pi/4 - \varphi_{{\bm q}}$ in the term proportional to the sines in the last line of Eq.~\ref{eq:Boltzmann_lin_rhs_2}, after some straighforward algebra we find that it can be rewritten as the term in the last-but-one line.
Therefore,
\be \label{eq:Boltzmann_lin_rhs_3}
{\cal D} u_{\mu\nu} 
&=& 
\frac{2\pi v_{\rm F}^2 \tau}{8 k_{\rm B} T} \sum_{{\bm q},{\bm k},{\bm k}'} \sum_{\lambda,\lambda',\lambda'',\lambda'''} |V_{\rm ee}({\bm q},\varepsilon_{{\bm k},\lambda} - \varepsilon_{{\bm k}+{\bm q},\lambda''})|^2 
{\cal F}({{\bm k},\lambda}; {\bm k}+{\bm q},\lambda'') {\cal F}({{\bm k}',\lambda'}; {\bm k}'-{\bm q},\lambda''') 
\nonumber\\
&\times&
\delta(\varepsilon_{{\bm k},\lambda}+\varepsilon_{{\bm k}',\lambda'}-\varepsilon_{{\bm k}+{\bm q},\lambda''}-\varepsilon_{{\bm k}'-{\bm q},\lambda'''})
f_{{\bm k},\lambda}^{(0)} f_{{\bm k}',\lambda'}^{(0)} (1-f_{{\bm k}+{\bm q},\lambda''}^{(0)})(1-f_{{\bm k}'-{\bm q},\lambda'''}^{(0)})
\nonumber\\
&\times&
\big[\lambda k \cos(2\varphi_{{\bm k}}) + \lambda' k' \cos(2\varphi_{{\bm k}'}) - \lambda'' |{\bm k}+{\bm q}|
\cos(2\varphi_{{\bm k}+{\bm q}}) - \lambda''' |{\bm k}' -{\bm q} |\cos(2\varphi_{{\bm k}'-{\bm q}}) \big]^2
u_{\mu\nu},
\ee
Which allows us to find 
\be \label{eq:Boltzmann_lin_tau}
\frac{1}{\tau_v}
&=& 
\frac{2\pi v_{\rm F}^2}{8 k_{\rm B} T {\cal D}} \sum_{{\bm q},{\bm k},{\bm k}'} \sum_{\lambda,\lambda',\lambda'',\lambda'''} |V_{\rm ee}({\bm q},\varepsilon_{{\bm k},\lambda} - \varepsilon_{{\bm k}+{\bm q},\lambda''})|^2 
{\cal F}({{\bm k},\lambda}; {\bm k}+{\bm q},\lambda'') {\cal F}({{\bm k}',\lambda'}; {\bm k}'-{\bm q},\lambda''') 
\nonumber\\
&\times&
\delta(\varepsilon_{{\bm k},\lambda}+\varepsilon_{{\bm k}',\lambda'}-\varepsilon_{{\bm k}+{\bm q},\lambda''}-\varepsilon_{{\bm k}'-{\bm q},\lambda'''})
f_{{\bm k},\lambda}^{(0)} f_{{\bm k}',\lambda'}^{(0)} (1-f_{{\bm k}+{\bm q},\lambda''}^{(0)})(1-f_{{\bm k}'-{\bm q},\lambda'''}^{(0)})
\nonumber\\
&\times&
\big[\lambda k \cos(2\varphi_{{\bm k}}) + \lambda' k' \cos(2\varphi_{{\bm k}'}) - \lambda'' |{\bm k}+{\bm q}|
\cos(2\varphi_{{\bm k}+{\bm q}}) - \lambda''' |{\bm k}' -{\bm q} |\cos(2\varphi_{{\bm k}'-{\bm q}}) \big]^2
.
\ee
Using that
\be \label{delta_split}
\delta(\varepsilon_{{\bm k},\lambda}+\varepsilon_{{\bm k}',\lambda'}-\varepsilon_{{\bm k}+{\bm q},\lambda''}-\varepsilon_{{\bm k}'-{\bm q},\lambda'''}) = 
\int_{-\infty}^\infty d\omega
\delta(\omega + \varepsilon_{{\bm k},\lambda} - \varepsilon_{{\bm k}+{\bm q},\lambda''})
\delta(\varepsilon_{{\bm k}',\lambda'}-\varepsilon_{{\bm k}'-{\bm q},\lambda'''} - \omega),
\ee
and that
\be
& f_{{\bm k},\lambda}^{(0)} (1-f_{{\bm k}+{\bm q},\lambda''}^{(0)}) = \big[1+n(\omega)\big] [f^{(0)}(\varepsilon_{{\bm k},\lambda}) - f^{(0)}(\varepsilon_{{\bm k},\lambda}+\omega)], \nn \\
& f_{{\bm k}',\lambda'}^{(0)}(1-f_{{\bm k}'-{\bm q},\lambda'''}^{(0)}) = -n(\omega) [f^{(0)}(\varepsilon_{{\bm k}',\lambda'}) - f^{(0)}(\varepsilon_{{\bm k}',\lambda'} - \omega)], \label{TDrelation}
\ee
we rewrite Eq.~\ref{eq:Boltzmann_lin_tau} as 
\be \label{eq:tau_v_def}
\frac{1}{\tau_v}
&=& 
-\frac{2\pi v_{\rm F}^2}{8 {\cal D}} \sum_{\lambda,\lambda',\lambda'',\lambda'''} 
\int_{-\infty}^\infty d\omega \int \frac{d^2 {\bm q}}{(2\pi)^2} \frac{|V_{\rm ee}({\bm q},\omega)|^2 }{4 k_{\rm B} T \sinh^2(\frac{\omega}{2k_{\rm B} T})}
\int \frac{d^2 {\bm k}}{(2\pi)^2}
\int \frac{d^2 {\bm k}'}{(2\pi)^2}
\delta(\omega + \varepsilon_{{\bm k},\lambda} - \varepsilon_{{\bm k}+{\bm q},\lambda''})
\nonumber\\
&\times&
\delta(\varepsilon_{{\bm k}',\lambda'}-\varepsilon_{{\bm k}'-{\bm q},\lambda'''} - \omega)
{\cal F}({{\bm k},\lambda};
{\bm k}+{\bm q},\lambda'') {\cal F}({{\bm k}',\lambda'}; {\bm k}'-{\bm q},\lambda''') 
[f^{(0)}(\varepsilon_{{\bm k},\lambda}) - f^{(0)}(\varepsilon_{{\bm k},\lambda}+\omega)] 
\nonumber\\
&\times&
[f^{(0)}(\varepsilon_{{\bm k}',\lambda'}) - f^{(0)}(\varepsilon_{{\bm k}',\lambda'} - \omega)]
\big[\lambda k \cos(2\varphi_{{\bm k}}) + \lambda' k' \cos(2\varphi_{{\bm k}'}) - \lambda'' |{\bm k}+{\bm q}|
\cos(2\varphi_{{\bm k}+{\bm q}}) - \lambda''' |{\bm k}' -{\bm q} |\cos(2\varphi_{{\bm k}'-{\bm q}}) \big]^2
.
\nn
\ee
In what follows we proceed to prepare this equation for numerical implementation.
\subsection{The integration boundaries}
We shift $\varphi_{\bm k}, \varphi_{{\bm k}'} \to \varphi_{\bm k} + \varphi_{\bm q}, \varphi_{{\bm k}'} + \varphi_{\bm q}$.
The delta function $\delta(\omega + \varepsilon_{{\bm k},\lambda} - \varepsilon_{{\bm k}+{\bm q},\lambda''})$ is solved by
\be
\lambda''(\omega + \lambda v_{\rm F} k) - v_{\rm F} \sqrt{k^2 + q^2 + 2 k q \cos(\varphi_{\bm k})} = 0
.
\ee
The solution of the delta function $\delta(\varepsilon_{{\bm k}',\lambda'}-\varepsilon_{{\bm k}'-{\bm q},\lambda'''} - \omega)$ can be found from the previous one by setting $k\to k', \lambda\to \lambda', \lambda''\to \lambda'''$ and $q,\omega \to -q,-\omega$.
Thus, we will not discuss it in detail, but directly quote the boundaries of integrations it produces at the end of this section.
The equation above has solution, when $\lambda''(\omega + \lambda v_{\rm F} k)>0$,
\be \label{eq:cos_phik_delta}
\cos(\varphi_{\bm k}^{(0)}) = \frac{\omega^2 + 2\lambda \omega v_{\rm F} k - v_{\rm F}^2 q^2}{2 v_{\rm F}^2 k q}.
\ee
Note that there are two solutions for $\varphi_{\bm k}^{(0)}$, one between $0$ and $\pi$ and one between $-\pi$ and $0$.
The integrand is symmetric in $\varphi_{\bm k}$, so one can consider only one of them and multiply the result by $2$.
The cosine must lie between $-1$ and $1$. Therefore, we must impose
\be \label{eq:delta_ineq}
\left\{
\begin{array}{l}
\lambda''(\omega + \lambda v_{\rm F} k) > 0
\vspace{0.2cm}\\
(\lambda \omega - v_{\rm F} q)(\lambda \omega + v_{\rm F} q + 2 v_{\rm F} k) < 0
\vspace{0.2cm}\\
(\lambda \omega + v_{\rm F} q)(\lambda \omega - v_{\rm F} q + 2 v_{\rm F} k) > 0
\end{array}
\right.
.
\ee
We can then distinguish $2$ cases for each of the last two equations
\be
\left\{
\begin{array}{l}
{\displaystyle q < \frac{\lambda \omega}{v_{\rm F}} }
\vspace{0.2cm}\\
{\displaystyle 0 < k < -\frac{\lambda \omega + v_{\rm F} q}{2 v_{\rm F}} }
\end{array}
\right.
\quad
\text{or}
\quad
\left\{
\begin{array}{l}
{\displaystyle q > \frac{\lambda \omega}{v_{\rm F}} }
\vspace{0.2cm}\\
{\displaystyle k > -\frac{\lambda \omega + v_{\rm F} q}{2 v_{\rm F}} }
\end{array}
\right.,
\ee
and
\be
\left\{
\begin{array}{l}
{\displaystyle q > - \frac{\lambda \omega}{v_{\rm F}} }
\vspace{0.2cm}\\
{\displaystyle k > \frac{v_{\rm F} q - \lambda \omega}{2 v_{\rm F}} }
\end{array}
\right.
\quad
\text{or}
\quad
\left\{
\begin{array}{l}
{\displaystyle q < - \frac{\lambda \omega}{v_{\rm F}} }
\vspace{0.2cm}\\
{\displaystyle 0< k < \frac{v_{\rm F} q - \lambda \omega}{2 v_{\rm F}} }
\end{array}
\right..
\ee
There are therefore $4$ cases ($\ell = 1, \ldots, 4$), which can be summarised as
\be
\left\{
\begin{array}{l}
{\displaystyle \max\left(0, q_{\rm min}^{(\ell)}(\lambda\omega) \right) < q < q_{\rm max}^{(\ell)}(\lambda\omega) }
\vspace{0.2cm}\\
{\displaystyle k_{\rm min}^{(\ell)}(q, \lambda\omega) < k < k_{\rm max}^{(\ell)}(q, \lambda\omega) }
\end{array}
\right.,
\ee
where
\be
\begin{array}{|c|c|c|c|c|}
\hline
 & \ell=1 & \ell=2 & \ell=3 & \ell=4
\\
\hline
q_{\rm min}^{(\ell)}(\lambda\omega) 
& 
{\displaystyle -\frac{\lambda \omega}{v_{\rm F}}} 
&
{\displaystyle 0 }
&
{\displaystyle \max\left(-\frac{\lambda \omega}{v_{\rm F}}, \frac{\lambda \omega}{v_{\rm F}} \right)}
& 
{\displaystyle \frac{\lambda \omega}{v_{\rm F}} }
\\
\hline
q_{\rm max}^{(\ell)}(\lambda\omega) 
& 
{\displaystyle \frac{\lambda \omega}{v_{\rm F}} }
&
{\displaystyle \min\left(-\frac{\lambda \omega}{v_{\rm F}}, \frac{\lambda \omega}{v_{\rm F}} \right)}
& 
{\displaystyle \infty}
& 
{\displaystyle 
-\frac{\lambda \omega}{v_{\rm F}} }
\\
\hline
k_{\rm min}^{(\ell)}(q, \lambda\omega) 
& 
{\displaystyle \max\left(0, k_0 \right)}
& 
0
&
{\displaystyle \max\left(k_0, k_1 \right)}
& 
{\displaystyle \max\left(0, k_1 \right)}
\\
\hline
k_{\rm max}^{(\ell)}(q, \lambda\omega) 
& 
{\displaystyle k_1 } 
&
{\displaystyle \min\left(k_0, k_1 \right)}
& 
{\displaystyle \infty}
& 
{\displaystyle k_0 } 
\\
\hline
\end{array}
\ee
Here we defined $k_0 = (v_{\rm F}q - \lambda \omega)/(2v_{\rm F})$ and $k_1 = -(v_{\rm F} q + \lambda \omega)/(2 v_{\rm F})$.
Obviously, one has to require that $q_{\rm max}^{(\ell)}(\lambda\omega) > 0$ and $k_{\rm max}^{(\ell)}(q,\lambda\omega)>0$, as well as account for the first inequality in Eq.~\ref{eq:delta_ineq}.
We find, for each of the cases above,
\be
\left\{
\begin{array}{l}
{\displaystyle \lambda'' = \lambda}
\vspace{0.2cm}\\
{\displaystyle \max\left(0, q_{\rm min}^{(\ell)}(\lambda\omega) \right) < q < q_{\rm max}^{(\ell)}(\lambda\omega) }
\vspace{0.2cm}\\
{\displaystyle \max\left( -\frac{\lambda \omega}{v_{\rm F}}, k_{\rm min}^{(\ell)}(q, \lambda\omega)\right) < k' < k_{\rm max}^{(\ell)}(q, \lambda\omega) }
\end{array}
\right.
\quad
\text{or}
\quad
\left\{
\begin{array}{l}
{\displaystyle \lambda'' = -\lambda}
\vspace{0.2cm}\\
{\displaystyle \max\left(0, q_{\rm min}^{(\ell)}(\lambda\omega) \right)< q < q_{\rm max}^{(\ell)}(\lambda\omega) }
\vspace{0.2cm}\\
{\displaystyle k_{\rm min}^{(\ell)}(q, \lambda\omega) < k' < \min\left(-\frac{\lambda \omega}{v_{\rm F}}, k_{\rm max}^{(\ell)}(q, \lambda\omega) \right) }
\end{array}
\right.
.
\ee

For the $\delta$-function $\delta(\varepsilon_{{\bm k}',\lambda'}-\varepsilon_{{\bm k}'-{\bm q},\lambda'''} - \omega)$ we find instead
\be \label{eq:cos_phik1_delta}
\cos(\varphi_{{\bm k}'}^{(0)}) = -\frac{\omega^2 - 2\lambda' \omega v_{\rm F} k' - v_{\rm F}^2 q^2}{2 v_{\rm F}^2 k' q},
\ee
and
\be
\left\{
\begin{array}{l}
{\displaystyle \lambda''' = \lambda'}
\vspace{0.2cm}\\
{\displaystyle \max\left(0, -q_{\rm max}^{(\ell)}(-\lambda'\omega)\right) < q <  -q_{\rm min}^{(\ell)}(-\lambda'\omega)}
\vspace{0.2cm}\\
{\displaystyle \max\left( \frac{\lambda' \omega}{v_{\rm F}}, k_{\rm min}^{(\ell)}(-q, -\lambda'\omega)\right) < k < k_{\rm max}^{(\ell)}(-q, -\lambda'\omega) }
\end{array}
\right.
\quad
\text{or}
\quad
\left\{
\begin{array}{l}
{\displaystyle \lambda''' = -\lambda'}
\vspace{0.2cm}\\
{\displaystyle \max\left(0, -q_{\rm max}^{(\ell)}(-\lambda'\omega)\right) < q < -q_{\rm min}^{(\ell)}(-\lambda'\omega) }
\vspace{0.2cm}\\
{\displaystyle k_{\rm min}^{(\ell)}(-q, -\lambda'\omega) < k < \min\left(\frac{\lambda' \omega}{v_{\rm F}}, k_{\rm max}^{(\ell)}(-q, -\lambda'\omega) \right) }
\end{array}
\right.
.
\nn
\ee

\subsection{The matrix element of the viscosity}
The matrix element on the last line of Eq.~\ref{eq:tau_v_def} reads
\be \label{eq:matr_el_1}
{\cal M}_{\lambda\lambda'}(k,k',\omega,q) &\equiv&
\big[\lambda k \cos(2\varphi_{{\bm k}}) + \lambda' k' \cos(2\varphi_{{\bm k}'}) - \lambda'' |{\bm k}+{\bm q}|
\cos(2\varphi_{{\bm k}+{\bm q}}) - \lambda''' |{\bm k}' -{\bm q} |\cos(2\varphi_{{\bm k}'-{\bm q}}) \big]^2
\nn \\
&=&
4 \big[\lambda k \cos^2(\varphi_{{\bm k}}) + \lambda' k' \cos^2(\varphi_{{\bm k}'}) - \lambda'' |{\bm k}+{\bm q}|
\cos^2(\varphi_{{\bm k}+{\bm q}}) - \lambda''' |{\bm k}' -{\bm q} |\cos^2(\varphi_{{\bm k}'-{\bm q}}) \big]^2
\nn \\
&=&
4 \left[\lambda k \cos^2(\varphi_{{\bm k}}) - \frac{\big[k \cos(\varphi_{{\bm k}}) + q \cos(\varphi_{{\bm q}})\big]^2}{\omega/v_{\rm F} + \lambda k}
+
\lambda' k' \cos^2(\varphi_{{\bm k}'})
- \frac{\big[k' \cos(\varphi_{{\bm k}'}) - q \cos(\varphi_{{\bm q}})\big]^2}{\lambda' k' - \omega/v_{\rm F}}
\right]^2
,
\nn
\ee
where we used the $\delta$-functions that impose energy conservation to cancel a term proportional to
\be
\lambda k + \lambda' k' - \lambda'' |{\bm k}+{\bm q}|
- \lambda''' |{\bm k}' -{\bm q}| =0,
\ee
and to replace 
\be
&&
\lambda'' |{\bm k}+{\bm q}|
\to \omega/v_{\rm F} + \lambda k,
\nn \\
&&
\lambda''' |{\bm k}'-{\bm q}|  \to \lambda' k' - \omega/v_{\rm F}.
\ee
If we shift $\varphi_{\bm k}, \varphi_{{\bm k}'} \to \varphi_{\bm k} + \varphi_{\bm q}, \varphi_{{\bm k}'} + \varphi_{\bm q}$, the matrix element in Eq.~\ref{eq:matr_el_1} is the only term that depends on $\varphi_{\bm q}$.
We can then expand sines and cosines, as well as the brackets, and perform the integration.
Finally, all odd terms in either $\varphi_{\bm k}$ or $\varphi_{{\bm k}'}$ vanish under angular integration, and we can use Eqs.~\ref{eq:cos_phik_delta} and~\ref{eq:cos_phik1_delta} to replace the remaining trigonometric functions.
The final result is quite lengthy and is not reported here.
\subsection{The integral of the delta function}
We now evaluate
\be
{\cal I}_\lambda(k,\omega,q) &=& 
\int_0^{2\pi} \frac{d \varphi_{\bm k}}{2\pi} \delta(\omega + \varepsilon_{{\bm k},\lambda} - \varepsilon_{{\bm k}+{\bm q},\lambda''}) {\cal F}({{\bm k},\lambda}; {\bm k}+{\bm q},\lambda'')
\nn \\
&=& 
\int_0^{\pi} \frac{d \varphi_{\bm k}}{2\pi} \delta(\omega + \varepsilon_{{\bm k},\lambda} - \varepsilon_{{\bm k}+{\bm q},\lambda''}) \left(1+ \lambda \frac{k + q \cos(\varphi_{\bm k} - \varphi_{\bm q})}{\lambda k + \omega/v_{\rm F}} \right)
.
\ee
We shift $\varphi_{\bm k}\to \varphi_{\bm k} + \varphi_{\bm q}$, and using Eq.~\ref{eq:cos_phik_delta} we get
\be
{\cal I}_\lambda(k,\omega,q) &=& 
\frac{2 \lambda \big[(\omega +2\lambda v_{\rm F} k)^2 - v_{\rm F}^2 q^2\big]}{2 v_{\rm F} k (\lambda v_{\rm F} k + \omega)} \frac{1}{2\pi} 
\frac{|\omega + \lambda v_{\rm F} k|}{v_{\rm F}^2 k q \sqrt{1 - \cos^2(\varphi_{\bm k}^{(0)})}}
.
\ee
Finally, using Eq.~\ref{eq:cos_phik_delta} we get
\be \label{eq:delta_func_int}
{\cal I}_\lambda(k,\omega,q)  = 
\frac{1}{2\pi} \frac{2 \lambda \big[(\omega +2\lambda v_{\rm F} k)^2 - v_{\rm F}^2 q^2\big]}{2 v_{\rm F} k (\lambda v_{\rm F} k + \omega)}
\frac{2|\omega + \lambda v_{\rm F} k|}{\sqrt{(2 v_{\rm F}^2 k q)^2 - (\omega^2 + 2\lambda \omega v_{\rm F} k - v_{\rm F}^2 q^2)^2}}
.
\ee
Similarly, we have 
\be \label{eq:delta_func_int_prime}
{\cal I}'_{\lambda'}(k',\omega,q) &=&
\int \frac{d \varphi_{\bm k}'}{2\pi}
\delta(\varepsilon_{{\bm k}',\lambda'}-\varepsilon_{{\bm k}'-{\bm q},\lambda'''} - \omega)
{\cal F}({{\bm k}',\lambda'}; {\bm k}'-{\bm q},\lambda''') 
\nn \\
&=&
\frac{1}{2\pi} \frac{2 \lambda' \big[(\omega - 2\lambda' v_{\rm F} k')^2 - v_{\rm F}^2 q^2\big]}{2 v_{\rm F} k' (\lambda' v_{\rm F} k' - \omega)}
\frac{2|\omega - \lambda' v_{\rm F} k'|}{\sqrt{(2 v_{\rm F}^2 k' q)^2 - (\omega^2 - 2\lambda' \omega v_{\rm F} k' - v_{\rm F}^2 q^2)^2}}
.
\ee

\subsection{Final integral for the viscosity transport time}
The final integral takes the form of Eq.~\ref{eq:viscosity_time_final}.
To find the temperature dependence of Eq.~\ref{eq:viscosity_time_final}, we scale the various quantities as follows 
\be
\omega = k_{\rm B} T \bar \omega, \quad q =  k_{\rm B} T \bar q/v_{\rm F}, \quad k = k_{\rm B} T \bar k/v_{\rm F}, \quad k' = k_{\rm B} T \bar k'/v_{\rm F}, \quad \mu = k_{\rm B} T \bar \mu.
\ee
The delta-function integrals are then 
\be
{\cal I}_\lambda(k,\omega,q)  &=& 
\frac{1}{2\pi} \frac{2\lambda (k_{\rm B} T)^2 \big[(\bar \omega +2\lambda \bar k)^2 - \bar q^2\big]}{(k_{\rm B} T)^2 2 \bar k (\lambda \bar k + \bar \omega)}
\frac{2 k_{\rm B} T |\bar \omega + \lambda \bar k|}{(k_{\rm B} T)^2 \sqrt{(2 \bar k \bar q)^2 - (\bar \omega^2 + 2\lambda \bar \omega \bar k -  \bar q^2)^2}} \nonumber \\
&=& 
\frac{1}{k_{\rm B} T}\bar {\cal I}_\lambda(\bar k,\bar \omega,\bar q),
\ee
and same for ${\cal I}'_{\lambda'}(k',\omega,q) = (k_{\rm B} T)^{-1} \bar {\cal I}'_{\lambda'}(\bar k',\bar \omega,\bar q)$.

The matrix element \eqref{eq:matr_el_1} scales as 
\be
{\cal M}_{\lambda\lambda'}(k,k',\omega,q) = (k_{\rm B} T)^2 \bar {\cal M}_{\lambda\lambda'}(\bar k,\bar k',\bar \omega,\bar q)
.
\ee

${\cal D}$ from \eqref{eq:D_def} is a function of $\mu$ and scales the following way
\be 
{\cal D(\mu)} 
&=& (k_{\rm B} T)^3 \cal \bar D(\bar \mu)
.
\ee
The electron-electron potential $V_{\rm ee}$ contains the dielectric function $\epsilon({\bm q},\omega) = 1 - v_{\bm q}\chi_{nn}({\bm q},\omega)$.
The density-density response can be estimated as 
\be
\Re e \chi^{(0)}_{nn}({\bm q},\omega) = -\frac{q^2}{16}\frac{\theta(v_{\rm F}^2 q^2 -\omega ^2)}{\sqrt{v_{\rm F}^2q^2 - \omega^2}} = -k_{\rm B} T \frac{\bar q^2}{16}\frac{\theta(\bar q^2 -\bar \omega ^2)}{\sqrt{\bar q^2 - \bar \omega^2}}=k_{\rm B} T \Re e \bar \chi^{(0)}_{nn}({\bm \bar q},\bar \omega).
\ee
Therefore, the electron-electron potential scales as 
\be
|V_{\rm ee}(q,\omega)|^2 = \frac{1}{(k_{\rm B} T)^2}|\bar V_{\rm ee}(\bar q,\bar \omega)|^2,
\ee
where ${\bar V}_{\rm ee}(\bar q,\bar \omega) = 
{\bar v}_{\bm q}/\epsilon({\bm q},\omega)$ with ${\bar v}_{\bm q} = 2\pi \alpha_{\rm ee}/\bar q$.
Here, $\alpha_{\rm ee} = e^2/(\hbar v_{\rm F} \kappa)$ is the graphene's fine structure constant.
The final expression for the viscosity transport time is 
\be 
\frac{1}{\tau_v}
&=& 
k_{\rm B} T C_v
,
\ee
where $C_v$ is a dimensionless quantity defined as the integral
\be  \label{eq:Cv_def}
C_v
&=& 
-\frac{2\pi}{8 {\cal \bar D}} \sum_{\lambda,\lambda',\lambda'',\lambda'''} 
\int_{-\infty}^\infty d\bar \omega 
\int_{}^{} d\bar q
\int_{}^{} d\bar k
\int_{}^{} d\bar k'
\frac{\bar q}{2\pi} 
\frac{\bar k}{2\pi}
\frac{\bar k'}{2\pi}
\frac{|V_{\rm ee}(\bar q,\bar \omega)|^2}{4 \sinh^2\left( \frac{\bar \omega}{2} \right) }
\bar {\cal I}_{\lambda}(\bar k,\bar \omega,\bar q)
\bar {\cal I}'_{\lambda'}(\bar k',\bar \omega,\bar q)
\nonumber\\
&\times&
[f^{(0)}(\varepsilon_{{\bm \bar k},\lambda}) - f^{(0)}(\varepsilon_{{\bm \bar k},\lambda}+\bar \omega)] 
[f^{(0)}(\varepsilon_{{\bm \bar k}',\lambda'}) - f^{(0)}(\varepsilon_{{\bm \bar k}',\lambda'} - \bar \omega)]
\bar {\cal M}_{\lambda\lambda'}(\bar k,\bar k',\bar \omega,\bar q)
.
\ee

\section{Viscosity in a parabolic-band 2DEG}\label{app:2deg}
The single-particle Hamiltonian introduced in Sec.~IV.B is
\begin{equation}\label{eq:2deg_hamiltonian_app}
    H = \frac{k^2}{2m}.
\end{equation}
The corresponding eigenstates and eigenvalues are
\be
    \ket{\bm k} = e^{i{\bm k}\cdot{\bm r}},
    \qquad
    \varepsilon_{\bm k} = \frac{k^2}{2m}.
\ee

Starting from the general kinetic framework of Sec.~\ref{sec:graphene viscosity}, we specialize Eqs.~\eqref{eq:boltzmann_def} and \eqref{eq:I_ee_def} to a single parabolic band.
{Because the form factors are unity for a 2DEG, the scattering kernel entering Eq.~\eqref{eq:I_ee_def} reduces to}
\bea
    W({\bm k},{\bm k}',{\bm q})
    &=&
    2\pi \left|V_{\rm ee}\big({\bm q},\varepsilon_{\bm k}-\varepsilon_{\bm k+\bm q}\big)\right|^2 .
    \label{eq:W_2deg}
\eea

The collision integral therefore becomes
\begin{eqnarray}\label{eq:Iee_2deg}
{\cal I}_{\rm ee}[f_{\bm k}({\bm r})]
&=&
\sum_{{\bm q},{\bm k}'}
W({\bm k},{\bm k}',{\bm q})
\delta(\varepsilon_{\bm k}+\varepsilon_{{\bm k}'}-\varepsilon_{{\bm k}+{\bm q}}-\varepsilon_{{\bm k}'-{\bm q}})
\nonumber\\
&\times&
\Big[
f_{\bm k} f_{{\bm k}'}
(1-f_{{\bm k}+{\bm q}})
(1-f_{{\bm k}'-{\bm q}})
-
(1-f_{\bm k})
(1-f_{{\bm k}'})
f_{{\bm k}+{\bm q}}
f_{{\bm k}'-{\bm q}}
\Big].
\end{eqnarray}

As in the main text, we write
\be
    f_{\bm k}({\bm r}) = f^{(0)}_{\bm k}({\bm r}) + \delta f_{\bm k}({\bm r}).
\ee
Linearizing the left-hand side of the Boltzmann equation in the velocity gradients gives
\bea
    {\cal L}_{\bm k}
    &=&
    v_{\bm k}^\alpha \partial_\alpha f^{(0)}_{\bm k}
    =
    \frac{k^\alpha k^\beta}{2m}
    \left(-\frac{\partial f^{(0)}_{\bm k}}{\partial \varepsilon_{\bm k}}\right)
    (\partial_\alpha u_\beta + \partial_\beta u_\alpha).
\eea
For incompressible flow, $\partial_\alpha u_\alpha = 0$, this becomes
\bea
    {\cal L}_{\bm k}
    &=&
    \left(-\frac{\partial f^{(0)}_{\bm k}}{\partial \varepsilon_{\bm k}}\right)
    A_{\bm k}^{\alpha\beta} u_{\alpha\beta},
    \label{eq:L_2deg}
\eea
where
\bea
    A_{\bm k}^{\alpha\beta}
    &=&
    \frac{k^2}{2m}
    \left(
    \sigma_z^{\alpha\beta}\cos 2\phi_{\bm k}
    +
    \sigma_x^{\alpha\beta}\sin 2\phi_{\bm k}
    \right),
    \\
    u_{\alpha\beta}
    &=&
    \frac{1}{2}
    \left(
    \partial_\alpha u_\beta
    +
    \partial_\beta u_\alpha
    -
    \delta_{\alpha\beta}\partial_\rho u_\rho
    \right).
\eea
We use the same single-mode ansatz as in the main text,
\be
    \delta f_{\bm k}
    =
    \tau_{\rm p}
    \left(-\frac{\partial f^{(0)}_{\bm k}}{\partial \varepsilon_{\bm k}}\right)
    A_{\bm k}^{\alpha\beta} u_{\alpha\beta}.
\ee

Projecting the linearized kinetic equation onto the shear tensor structure is cleaner if the left- and right-hand sides are treated separately.
We therefore define
\bea
    I_{\rm LHS}^{\mu\nu}
    &\equiv&
    \sum_{\bm k}
    \left(-\frac{\partial f^{(0)}_{\bm k}}{\partial \varepsilon_{\bm k}}\right)
    A_{\bm k}^{\mu\nu}
    A_{\bm k}^{\alpha\beta}
    u_{\alpha\beta},
    \label{eq:ILHS_2deg}
    \\
    I_{\rm RHS}^{\mu\nu}
    &\equiv&
    \frac{2\pi \tau_{\rm p}}{4 k_{\rm B} T}
    \sum_{{\bm q},{\bm k},{\bm k}'}
    \left|V_{\rm ee}\big({\bm q},\varepsilon_{\bm k}-\varepsilon_{{\bm k}+{\bm q}}\big)\right|^2
    \delta(\varepsilon_{\bm k}+\varepsilon_{{\bm k}'}-\varepsilon_{{\bm k}+{\bm q}}-\varepsilon_{{\bm k}'-{\bm q}})
    \nonumber\\
    &&\times
    f_{\bm k} f_{{\bm k}'}
    (1-f_{{\bm k}+{\bm q}})
    (1-f_{{\bm k}'-{\bm q}})
    \Big[
    A_{\bm k}^{\alpha\beta}
    +
    A_{{\bm k}'}^{\alpha\beta}
    -
    A_{{\bm k}+{\bm q}}^{\alpha\beta}
    -
    A_{{\bm k}'-{\bm q}}^{\alpha\beta}
    \Big]
    \nonumber\\
    &&\times
    \Big[
    A_{\bm k}^{\mu\nu}
    +
    A_{{\bm k}'}^{\mu\nu}
    -
    A_{{\bm k}+{\bm q}}^{\mu\nu}
    -
    A_{{\bm k}'-{\bm q}}^{\mu\nu}
    \Big]
    u_{\alpha\beta}.
    \label{eq:IRHS_2deg_def}
\eea

The angular average on the left-hand side satisfies
\bea
    \int \frac{d\phi_{\bm k}}{2\pi}
    A_{\bm k}^{\mu\nu}
    A_{\bm k}^{\alpha\beta}
    u_{\alpha\beta}
    &=&
    \left(\frac{k^2}{2m}\right)^2 u_{\mu\nu}.
    \label{eq:A_identity_2deg}
\eea
Using Eq.~\eqref{eq:A_identity_2deg}, Eq.~\eqref{eq:ILHS_2deg} becomes
\bea
    I_{\rm LHS}^{\mu\nu}
    &=&
    {\cal D}_{\rm p} u_{\mu\nu},
    \label{eq:ILHS_eval_2deg}
\eea
where
\bea
    {\cal D}_{\rm p}(\mu,T)
    &=&
    \int \frac{d\phi\,dk}{(2\pi)^2}
    k
    \left(\frac{k^2}{2m}\right)^2
    \left(-\frac{\partial f^{(0)}(\varepsilon_{\bm k}-\mu)}{\partial \varepsilon_{\bm k}}\right)
    \nonumber\\
    &=&
    - \frac{N m (k_{\rm B}T)^2}{4\pi}
    {\rm Li}_2\!\left(-e^{\mu/k_{\rm B}T}\right).
    \label{eq:Dp_appendix}
\eea

After symmetrizing Eq.~\eqref{eq:IRHS_2deg_def} in the standard way, the right-hand side can be written as
\bea
    I_{\rm RHS}^{\mu\nu}
    &=&
    \frac{2\pi \tau_{\rm p}}{8 m^2 k_{\rm B} T}
    \sum_{{\bm q},{\bm k},{\bm k}'}
    \left|V_{\rm ee}\big({\bm q},\varepsilon_{\bm k}-\varepsilon_{{\bm k}+{\bm q}}\big)\right|^2
    \delta(\varepsilon_{\bm k}+\varepsilon_{{\bm k}'}-\varepsilon_{{\bm k}+{\bm q}}-\varepsilon_{{\bm k}'-{\bm q}})
    \nonumber\\
    &&\times
    f_{\bm k} f_{{\bm k}'}
    (1-f_{{\bm k}+{\bm q}})
    (1-f_{{\bm k}'-{\bm q}})
    {\cal M}_{\rm p}({\bm k},{\bm k}',{\bm q})
    u_{\mu\nu},
    \label{eq:IRHS_2deg}
\eea
where
\bea
    {\cal M}_{\rm p}({\bm k},{\bm k}',{\bm q})
    &=&
    \Big[
    k^2\cos(2\phi_{\bm k})
    +
    k'^2\cos(2\phi_{{\bm k}'})
    -
    ({\bm k}'-{\bm q})^2\cos(2\phi_{{\bm k}'-{\bm q}})
    -
    ({\bm k}+{\bm q})^2\cos(2\phi_{{\bm k}+{\bm q}})
    \Big]^2
    \nonumber\\
    &=&
    16 q^2 \cos^2\phi_{\bm q}
    \Big[
    q\cos\phi_{\bm q}
    +
    k\cos\phi_{\bm k}
    -
    k'\cos\phi_{{\bm k}'}
    \Big]^2.
    \label{eq:Mp_intermediate}
\eea

To evaluate the angular delta-functions, we define
\bea
    {\cal I}_{\rm p}(k,q,\omega)
    &\equiv&
    \int \frac{d \phi_{\bm k}}{2\pi}
    \delta(\omega+\varepsilon_{\bm k}-\varepsilon_{{\bm k}+{\bm q}}).
    \label{eq:Ip_def_appendix}
\eea
The argument of the delta-function vanishes when
\bea
    \omega
    -
    \frac{q^2}{2m}
    -
    \frac{k q}{m}\cos(\phi_{\bm k}-\phi_{\bm q})
    &=&
    0.
    \label{eq:delta_condition_2deg}
\eea
After shifting $\phi_{\bm k}\to\phi_{\bm k}-\phi_{\bm q}$, Eq.~\eqref{eq:delta_condition_2deg} gives
\be
    \cos \phi_{\bm k}^{(0)}
    =
    \frac{\omega-q^2/(2m)}{kq/m},
    \qquad
    k
    \geq
    \left|\frac{m\omega}{q}-\frac{q}{2}\right|.
    \label{eq:phi_solution_2deg}
\ee
Therefore,
\bea
    {\cal I}_{\rm p}(k,q,\omega)
    &=&
    \frac{2m}{\pi}
    \frac{\theta\!\left(k-\left|\frac{q^2-2m\omega}{2q}\right|\right)}
    {\sqrt{4k^2q^2-(q^2-2m\omega)^2}},
    \nonumber\\
    {\cal I}_{\rm p}(k',q,-\omega)
    &=&
    \frac{2m}{\pi}
    \frac{\theta\!\left(k'-\left|\frac{q^2+2m\omega}{2q}\right|\right)}
    {\sqrt{4k'^2q^2-(q^2+2m\omega)^2}}.
    \label{eq:Ip_appendix}
\eea

Using Eq.~\eqref{eq:phi_solution_2deg} in Eq.~\eqref{eq:Mp_intermediate} and then averaging over $\phi_{\bm q}$, we obtain
\be
    {\cal M}_{\rm p}(k,k',q,\omega)
    =
    2 ( k^2 + k'^2 ) q^2 - q^4 - 4 m^2 \omega^2.
    \label{eq:Mp_final_appendix}
\ee

Equating Eq.~\eqref{eq:ILHS_eval_2deg} and Eq.~\eqref{eq:IRHS_2deg}, and then using Eq.~\eqref{eq:Ip_appendix} together with Eq.~\eqref{eq:Mp_final_appendix}, yields
\bea
    \frac{1}{\tau_{\rm p}}
    &=&
    \frac{-2\pi}{8 m^2 k_{\rm B} T {\cal D}_{\rm p}}
    \int dq\,dk\,dk'\,d\omega
    \frac{q}{2\pi}
    \frac{k}{2\pi}
    \frac{k'}{2\pi}
    \frac{|V_{\rm ee}({\bm q},\omega)|^2}{4 \sinh^2 \left( \frac{\omega}{2 k_{\rm B}T} \right)}
    \nonumber\\
    &&\times
    \left[
    f^{(0)}(\varepsilon_{\bm k}) - f^{(0)}(\varepsilon_{\bm k}+\omega)
    \right]
    \left[
    f^{(0)}(\varepsilon_{\bm k'}) - f^{(0)}(\varepsilon_{\bm k'}-\omega)
    \right]
    {\cal I}_{\rm p}(k,q,\omega)
    {\cal I}_{\rm p}(k',q,-\omega)
    {\cal M}_{\rm p}(k,k',q,\omega).
    \label{eq:tau_p_appendix}
\eea
Equation~\eqref{eq:tau_p_appendix} is the form of Eq.~\eqref{eq:tau_parabolic} in the main text.

With $\hbar=1$, the thermal wavevector is $k_T=\sqrt{2m k_{\rm B}T}$.
Introducing dimensionless variables,
\be
    k=\bar{k}k_T,
    \qquad
    k'=\bar{k}'k_T,
    \qquad
    q=\bar{q}k_T,
    \qquad
    \omega=\bar{\omega}k_{\rm B}T,
    \qquad
    \mu=\bar{\mu}k_{\rm B}T,
\ee
we have
\bea
    {\cal M}_{\rm p}
    &=&
    k_T^4 \bar{\cal M}_{\rm p},
    \qquad
    {\cal I}_{\rm p}
    =
    \frac{1}{k_{\rm B} T}\bar{\cal I}_{\rm p},
    \qquad
    {\cal D}_{\rm p}
    =
    (k_{\rm B} T)^2 \bar{\cal D}_{\rm p},
    \\
    \bar{\cal D}_{\rm p}
    &=&
    - \frac{N m}{4\pi}
    {\rm Li}_2\!\left(-e^{\bar{\mu}}\right).
\eea

For completeness, the particle density is related to the chemical potential by
\bea
    {\rm DoS}(\varepsilon)
    &=&
    \frac{N m}{2\pi},
    \\
    n
    &=&
    \int_0^\infty d\varepsilon\,
    {\rm DoS}(\varepsilon)\,
    f^{(0)}(\varepsilon-\mu)
    =
    \frac{N m}{2\pi}
    k_{\rm B} T
    \ln\!\left(1+e^{\mu/k_{\rm B}T}\right),
    \\
    \mu(T,n)
    &=&
    k_{\rm B} T
    \ln\!\left[
    \exp\!\left(\frac{2\pi n}{N m k_{\rm B}T}\right)-1
    \right].
\eea

\section{2DEG in the doped regime}\label{app:2deg_doped}
In the doped regime, Eq.~\eqref{eq:tau_p_appendix} is dominated by momenta on the Fermi surface,
\be
    |\bm k|
    =
    |\bm k'|
    =
    |\bm k+\bm q|
    =
    |\bm k'-\bm q|
    =
    k_F.
\ee
The differences of Fermi functions become
\bea
    f^{(0)}(\varepsilon_{\bm k}) - f^{(0)}(\varepsilon_{\bm k}+\omega)
    &\simeq&
    -\omega \frac{\partial f^{(0)}_{\bm k}}{\partial \varepsilon_{\bm k}}
    =
    \frac{m\omega}{k_F}\delta(k-k_F),
    \nonumber\\
    f^{(0)}(\varepsilon_{\bm k'}) - f^{(0)}(\varepsilon_{\bm k'}-\omega)
    &\simeq&
    \omega \frac{\partial f^{(0)}_{\bm k'}}{\partial \varepsilon_{\bm k'}}
    =
    -\frac{m\omega}{k_F}\delta(k'-k_F).
    \label{eq:fermi_surface_2deg}
\eea
The Fermi-surface constraints also imply
\be
    \phi_{\bm k} = \pi + \phi_{\bm k'},
    \qquad
    \phi_{\bm k+\bm q} = \pi + \phi_{\bm k'-\bm q},
\ee
so that $q \leq 2k_F$.
Using Eq.~\eqref{eq:Mp_final_appendix} at $k=k'=k_F$ and $\omega \to 0$, we find
\be
    {\cal M}_{\rm p}(k_F,k_F,q,0)
    =
    4k_F^2 q^2 - q^4.
    \label{eq:Mp_doped_2deg}
\ee
Likewise, Eq.~\eqref{eq:Ip_appendix} gives
\be
    {\cal I}_{\rm p}(k_F,q,0)\,
    {\cal I}_{\rm p}(k_F,q,0)
    =
    \left(\frac{2m}{\pi}\right)^2
    \frac{\theta(2k_F-q)}{4k_F^2 q^2-q^4}.
    \label{eq:Ip_doped_2deg}
\ee
The remaining frequency integral is
\be
    \int_{-\infty}^{\infty}d\omega\,
    \frac{\omega^2}{4\sinh^2(\omega/2k_{\rm B} T)}
    =
    \frac{2\pi^2}{3}(k_{\rm B} T)^3.
    \label{eq:omega_integral_2deg}
\ee

Substituting Eq.~\eqref{eq:fermi_surface_2deg}, Eq.~\eqref{eq:Mp_doped_2deg}, Eq.~\eqref{eq:Ip_doped_2deg}, and Eq.~\eqref{eq:omega_integral_2deg} into Eq.~\eqref{eq:tau_p_appendix} gives
\be
    \frac{1}{\tau_{\rm p}}
    =
    \frac{2 \pi m (k_{\rm B} T)^2}{3 (2\pi)^2\varepsilon_F^2}
    \left(\frac{2m}{\pi}\right)^2
    \int_0^{2k_F} dq\, q\, |V_{\rm ee}({\bm q},0)|^2.
    \label{eq:taup_doped_bare}
\ee
For the bare Coulomb interaction this integral diverges at the lower limit, so screening is essential.
Following the static-RPA discussion in Sec.~IV.B, we use
\be
    V_{\rm ee}^{\rm scr}(q,0)
    =
    \frac{2\pi \alpha}{q+q_{\rm TF}},
    \qquad
    \alpha = \frac{e^2}{\kappa},
    \qquad
    q_{\rm TF} = 2\pi e^2 N(0) = \frac{16 m \alpha}{\pi}.
    \label{eq:screened_potential_2deg}
\ee
The $q$ integral then becomes
\be
    \int_0^{2k_F} dq\, q\, \left|V_{\rm ee}^{\rm scr}(q,0)\right|^2
    =
    (2\pi\alpha)^2
    \left[
    -\frac{2k_F}{2k_F+q_{\rm TF}}
    -
    \ln\!\left(\frac{q_{\rm TF}}{q_{\rm TF}+2k_F}\right)
    \right].
    \label{eq:q_integral_2deg}
\ee
In the same limit, Eq.~\eqref{eq:Dp_appendix} reduces to
\be
    {\cal D}_{\rm p}
    \rightarrow
    \frac{m}{2\pi}\varepsilon_F^2.
    \label{eq:Dp_doped_2deg}
\ee
Therefore, the expression for the collision rate reads
\be
    \frac{1}{\tau_{\rm p}}
    =
    \frac{2 \pi \alpha^2}{3}
    \frac{m}{\varepsilon_F}
    \frac{(k_{\rm B} T)^2}{\varepsilon_F}
    \int_0^{2k_F} \frac{q\,dq}{(q+q_{\rm TF})^2}.
    \label{eq:taup_doped_final_appendix}
\ee

\section{Valley-polarized collision rate}\label{app:valleys}
{To match Sec.~\ref{sec:valley_polarized}, the explicit index $\Lambda=0,1$ labels the two displaced Dirac cones within a mini-Brillouin zone, while the generic label $\lambda$ retained from App.~\ref{app:collintder} should be read here as the band index $s=\pm1$. We assume these two cones to be sufficiently far apart in momentum space that exchange between them is negligible. Let their centers be separated by $\vec{G}$, so that our approximation requires $|\vec{k}| \ll |\vec{G}|$. Therefore, the initial-final state overlaps ${\cal F}({\bm k},\lambda, \Lambda; {\bm k}', \lambda', \Lambda)=\big|\langle {\bm k}, \lambda, \Lambda | {\bm k}', \lambda', \Lambda \rangle \big|^2={\cal F}({\bm k},\lambda; {\bm k}', \lambda')$ do not depend on the particular cone. The same holds for the angular phase-space factors in Eqs.~\eqref{eq:delta_func_int} and \eqref{eq:delta_func_int_prime}, and for the matrix element ${\cal M}$. The only change is therefore in the arguments of the distribution functions.}
\be  \label{eq:viscosity_time_valley}
\frac{1}{\tau_v}
&=& 
-\frac{2\pi v_{\rm F}^2}{8 {\cal D}} \sum_{\lambda,\lambda',\lambda'',\lambda'''} 
\int_{-\infty}^\infty d\omega 
\int_{}^{} dq
\int_{}^{} dk
\int_{}^{} dk'
\frac{q}{2\pi} \frac{|V_{\rm ee}(q,\omega)|^2}{4 k_{\rm B} T \sinh^2(\frac{\omega}{2k_{\rm B} T})}
\frac{k}{2\pi}
\frac{k'}{2\pi}
{\cal I}_{\lambda}(k,\omega,q)
{\cal I}'_{\lambda'}(k',\omega,q)
\nonumber\\
&\times&
\sum_{\Lambda, \Lambda'}[f^{(0)}(\varepsilon_{{\bm k},\lambda, \Lambda}) - f^{(0)}(\varepsilon_{{\bm k},\lambda, \Lambda}+\omega)] 
[f^{(0)}(\varepsilon_{{\bm k}',\lambda', \Lambda'}) - f^{(0)}(\varepsilon_{{\bm k}',\lambda', \Lambda'} - \omega)]
{\cal M}_{\lambda\lambda'}(k,k',\omega,q)
,
\ee

\end{widetext}

\bibliographystyle{apsrev4-2}
\bibliography{refs_populated}

\end{document}